\pdfoutput = 1
\documentclass[12pt,a4paper]{article}

\usepackage[top=1in, bottom=1.25in, left=1in, right=1in]{geometry}

\usepackage{amsmath}
\usepackage[dvipsnames]{xcolor}
\usepackage{graphicx}
\usepackage{booktabs}
\usepackage[colorlinks=true, allcolors=Blue]{hyperref}

\usepackage{xspace}
\newcommand{\BdJpsiK}{\ensuremath{B_d^0\to J/\psi K^0}\xspace}
\newcommand{\BdJpsiKS}{\ensuremath{B_d^0\to J/\psi K^0_{\text{S}}}\xspace}
\newcommand{\BdJpsiKL}{\ensuremath{B_d^0\to J/\psi K^0_{\text{L}}}\xspace}
\newcommand{\BsJpsiKS}{\ensuremath{B_s^0\to J/\psi K^0_{\text{S}}}\xspace}
\newcommand{\BdJpsiPi}{\ensuremath{B_d^0\to J/\psi\pi^0}\xspace}
\newcommand{\BdJpsiPhi}{\ensuremath{B_d^0\to J/\psi\phi}\xspace}
\newcommand{\BsJpsiPhi}{\ensuremath{B_s^0\to J/\psi\phi}\xspace}
\newcommand{\BdJpsiRho}{\ensuremath{B_d^0\to J/\psi\rho^0}\xspace}
\newcommand{\BJpsiX}{\ensuremath{B_q^0\to J/\psi X}\xspace}
\newcommand{\BJpsiP}{\ensuremath{B_q^0\to J/\psi P}\xspace}
\newcommand{\Bdpilnu}{\ensuremath{B_d^0\to\pi^-\ell^+\nu_{\ell}}\xspace}
\newcommand{\BsKlnu}{\ensuremath{B_s^0\to K^-\ell^+\nu_{\ell}}\xspace}

\makeatletter
\g@addto@macro\bfseries{\boldmath}
\makeatother

\usepackage{ifthen}
\newboolean{articletitles}
\setboolean{articletitles}{true}
\newboolean{inbibliography}
\setboolean{inbibliography}{false}
\usepackage{cite}
\usepackage{mciteplus}

\begin{document}

%
%
%
\begin{titlepage}
\vspace*{-0.7truecm}
\begin{flushright}
Nikhef-2020-036 \\
\end{flushright}

\vspace{1.6truecm}

\begin{center}
{\Large{\textbf{In Pursuit of New Physics with \BdJpsiK and \BsJpsiPhi Decays at the High-Precision Frontier}}}
\end{center}

\vspace{0.8truecm}

\begin{center}
{\bf Marten Z.\ Barel${}^{a}$, Kristof De Bruyn\,${}^{a,b}$, Robert Fleischer\,${}^{a,c}$ and\\ 
Eleftheria Malami\,${}^{a}$}

\vspace{0.5truecm}

${}^a${\sl Nikhef, Science Park 105, 1098 XG Amsterdam, Netherlands}

${}^b${\sl Van Swinderen Institute for Particle Physics and Gravity, University of Groningen, 9747 Groningen, Netherlands}

${}^c${\sl  Faculty of Science, Vrije Universiteit Amsterdam,\\
1081 HV Amsterdam, Netherlands}
\end{center}

\vspace*{1.7cm}

\begin{abstract}
\noindent
The decays \BdJpsiKS and \BsJpsiPhi play a key role for the determination of the $B^0_q$--$\bar B^0_q$ ($q=d,s$) mixing phases $\phi_d$ and $\phi_s$, respectively.
The theoretical precision of the extraction of these quantities is limited by doubly Cabibbo-suppressed penguin topologies, which can be included through control channels by means of the $SU(3)$ flavour symmetry of strong interactions.
Using the currently available data and a new simultaneous analysis, we discuss the state-of-the-art picture of these effects and include them in the extracted $\phi_q$ values.
We have a critical look at the Standard Model predictions of these phases and explore the room left for new physics.
Considering future scenarios for the high-precision era of flavour physics, we illustrate that we may obtain signals for physics beyond the Standard Model with a significance well above five standard deviations.
We also determine effective colour-suppression factors of \BdJpsiK, \BsJpsiKS and \BdJpsiPi decays, which serve as benchmarks for QCD calculations of the underlying decay dynamics, and present a new method using information from semileptonic \Bdpilnu and \BsKlnu decays.
\end{abstract}

\vfill

\noindent
January 2021

\end{titlepage}

\thispagestyle{empty}
\vbox{}
\newpage

\setcounter{page}{1}

%
%
%
\section{Introduction}
High precision measurements of the CP-violating phases $\phi_d$ and $\phi_s$, which are associated with the phenomenon of $B_q^0$--$\bar B_q^0$ mixing of the neutral $B_q$ mesons $(q=d,s)$, are part of the core physics programmes at the Large Hadron Collider (LHC) and the SuperKEKB accelerator, and will remain so for the next decades.
They offer excellent opportunities to search for evidence of New Physics (NP) processes that are not accounted for by the Standard Model (SM) paradigm.
In order to maximise the impact of these searches and fully exploit the future data, it is crucial to have a critical look at the theoretical SM interpretation of the underlying observables and to control the corresponding uncertainties, matching them with the experimental uncertainties.

The mixing phases $\phi_d$ and $\phi_s$ can be parametrised as
\begin{equation}\label{eq:B_mixing_phases}
    \phi_d \equiv  \phi_d^{\text{SM}} + \phi_d^{\text{NP}} = 2\beta + \phi_d^{\text{NP}}\:, \qquad
    \phi_s \equiv  \phi_s^{\text{SM}} + \phi_d^{\text{NP}}= -2\lambda^2\eta + \phi_s^{\text{NP}}\:,
\end{equation} 
where $\beta$ is one of the angles of the Unitarity Triangle (UT), and $\lambda$ and $\eta$ are two of the Wolfenstein parameters \cite{Wolfenstein:1983yz,Buras:1994ec} of the Cabibbo--Kobayashi--Maskawa (CKM) quark-mixing matrix \cite{Cabibbo:1963yz,Kobayashi:1973fv}.
The phases $\phi_q^{\text{NP}}$ describe contributions from potential new sources of CP violation lying beyond the SM.
To find such non-vanishing NP phases, we need to determine the phases $\phi_q$ and $\phi_q^{\text{SM}}$ as precisely as possible.
For the SM predictions, this requires a critical look at the input observables used in the UT fit, which we will briefly discuss in Section \ref{sec:SM} below.

The phases $\phi_d$ and $\phi_s$ are experimentally accessible through the charge--parity (CP) asymmetry arising from the interference of the $B_q^0$--$\bar B_q^0$ mixing process with the subsequent decays of the $B_q$ mesons into a CP eigenstate $f$.
The mixing-induced CP asymmetry allows us to measure an effective phase $\phi_{q,f}^{\text{eff}}$ which is given as follows:
\begin{equation}\label{eq:eff_mix_phase}
    \phi_{q,f}^{\text{eff}} \equiv \phi_q + \Delta\phi_q^f\:,
\end{equation}
where $\Delta\phi_q^f$ is a decay-channel-specific hadronic phase shift.
This relation is particularly favourable for \BdJpsiKS and \BsJpsiPhi decays, which are dominated by colour-suppressed tree amplitudes.
In case these topologies were the only contributions, we get $\Delta\phi_q^f = 0$.
Consequently, the effective phase $\phi_{q,f}^{\text{eff}}$ would equal the $B_q^0$--$\bar B_q^0$ mixing phase $\phi_q$, thereby allowing a direct measurement of this quantity.
However, these decays do also get contributions from doubly Cabibbo-suppressed penguin topologies, resulting in a shift $\Delta\phi_q^f$ of the order of $0.5^{\circ}$ \cite{Fleischer:1999nz,Fleischer:1999zi,Fleischer:1999sj,Ciuchini:2005mg,Faller:2008zc,Faller:2008gt,DeBruyn:2010hh,Ciuchini:2011kd,Jung:2012mp,Liu:2013nea,DeBruyn:2014oga,Frings:2015eva}.

In view of the current experimental precision, the absence of large NP effects and the prospects of the upgrade programmes at the LHC and SuperKEKB, this correction can no longer be considered as negligible.
A clear distinction between the experimental observable $\phi_{q,f}^{\text{eff}}$ and the theoretical parameter $\phi_q$ needs to be made in the interpretation of CP asymmetry measurements.
This is of particular importance when averaging these results with measurements from other decay channels.
The phase shift $\Delta\phi_q^f$ originates from non-perturbative, strong interaction effects, which depend on the dynamics of the decay in question.
The impact of the penguin topologies on the effective mixing phase, i.e.\ the size of $\Delta\phi_q^f$, is thus different for the various decay channels.
The average of the effective mixing phases therefore has no clear theoretical interpretation.
Instead, we must first correct all effective mixing phases individually, before making the average.
This paper focuses on the determination of the penguin corrections for \BdJpsiK and \BsJpsiPhi, while similar corrections have been discussed in Ref.~\cite{Fleischer:1999nz,Bel:2015wha} for $B_s^0\to D_s^+D_s^-$, in Ref.~\cite{Fleischer:2016jbf,Fleischer:2016ofb} for $B_s^0\to K^+K^-$, in Ref.~\cite{Fleischer:2011au} for $B_s^0\to J/\psi f_0(980)$, and in Ref.~\cite{Fleischer:2011ib} for $B_s^0\to J/\psi\eta^{(\prime)}$.

The phase shift $\Delta\phi_q^f$ cannot be calculated in a reliable way in QCD with the currently available theoretical methods.
Fortunately, applying the $SU(3)$ flavour symmetry of strong interactions, we may determine the impact of the doubly Cabibbo-suppressed penguin topologies through experimental data.
To this end, we relate the \BdJpsiKS and \BsJpsiPhi decays to partner control channels where the penguin contributions are not doubly Cabibbo-suppressed but enter in a Cabibbo-favoured way.
For the \mbox{\BdJpsiKS} decay, key control modes are \BsJpsiKS \cite{DeBruyn:2010hh,DeBruyn:2014oga} and \BdJpsiPi \cite{Fleischer:1999zi,Faller:2008zc}, while for the \BsJpsiPhi channel the main control mode is \BdJpsiRho \cite{Fleischer:1999zi,Faller:2008gt,DeBruyn:2014oga}.
The $B_s^0\to J/\psi \overline{K}^{*0}$ decay forms and alternative \cite{Faller:2008gt,DeBruyn:2014oga,Aaij:2015mea} to the \BdJpsiRho control mode, but its potential for the high-precision era is more limited as it is not a CP eigenstate and thus has no mixing-induced CP asymmetry to help constrain the penguin parameters.

We determine the penguin effects and their impact on the determination of $\phi_d$ and $\phi_s$ using the latest measurements of the \BdJpsiK, \BdJpsiPi, \BsJpsiKS, \BsJpsiPhi and \BdJpsiRho observables.
(Where measurements from \BdJpsiKS and \BdJpsiKL are combined, we will refer to these decays simply as \BdJpsiK.)
This analysis allows us to extract the values of $\phi_d$ and $\phi_s$, taking the hadronic penguin corrections into account.
To minimise the theoretical uncertainties associated with the breaking of the $SU(3)$-symmetry relations between these modes, we primarily focus on the information from the CP asymmetries to determine the penguin contributions.
Having the corresponding parameters at hand, we use the branching fraction measurements to study the dynamics of these decays.
We propose to utilise the branching fraction information provided by the differential rates of the semileptonic \Bdpilnu and \BsKlnu modes to extract the effective colour-suppression factors of the \BdJpsiPi and \mbox{\BsJpsiKS} decays without having to rely on the form factors.
These colour-suppression factors serve as benchmarks for QCD calculations of the underlying decay dynamics, and can also be used to test the $SU(3)$ flavour symmetry, which is a key input for our analysis.
These tests do not indicate large non-factorisable $SU(3)$-breaking corrections.

The outline of this paper is as follows:
Section \ref{sec:SM} briefly discusses our SM predictions for $\phi_d$ and $\phi_s$.
Section \ref{sec:PenFrame} introduces our formalism to deal with the penguin effects in the determination of the $B_q^0$--$B_q^0$ mixing phases, which we apply to the current experimental data in Section \ref{sec:CurrData}.
In Section \ref{sec:BR}, we combine the resulting information for the penguin parameters with branching fraction information to determine the hadronic parameters governing the \BJpsiX decays.
Here we propose a new strategy of adding information from semileptonic \Bdpilnu and \BsKlnu decays to the analysis.
In Section \ref{sec:bench}, we illustrate how the current picture may become much sharper as the experimental measurements are getting more precise in the future high-precision era of flavour physics.
Finally, we summarise our conclusions in Section \ref{sec:conclusion}

%
%
%
\section{Standard Model Predictions}\label{sec:SM}
\begin{figure}
    \centering
    \includegraphics[width=0.49\textwidth]{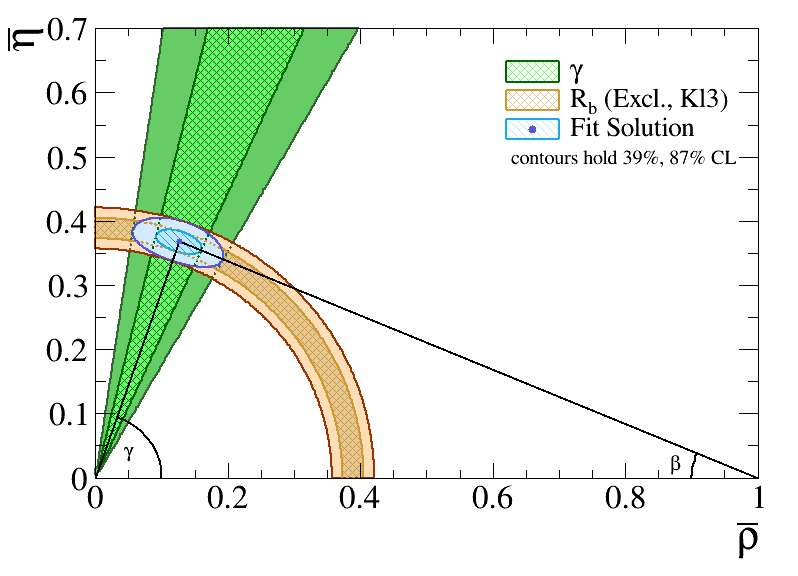}
    \caption{Determination of the UT apex $(\bar\rho,\bar\eta)$, where $\bar\rho \equiv \left(1-\lambda^2/2\right) \rho$ and \mbox{$\bar\eta \equiv \left(1-\lambda^2/2\right) \eta$,} from the measurements of the side $R_b$ and the angle $\gamma$, which can both be determined from decays with tree topologies only.}
    \label{fig:SM_Apex}
\end{figure}

The most accurate determination of the UT angle $\beta$ and the Wolfenstein parameters $\lambda$ and $\eta$, needed to calculate $\phi_d^{\text{SM}}$ and $\phi_s^{\text{SM}}$, comes from the global UT fits \cite{Charles:2015gya}.
However, we cannot blindly rely on these results because potential NP contributions can enter any of the observables used as input to these fits, and the results are often not independent from the experimental measurements of $\phi_d$ and $\phi_s$.
Instead, the most transparent approach to obtain SM predictions of the UT apex $(\bar\rho,\bar\eta)$, from which both $\phi_d^{\text{SM}}$ and $\phi_s^{\text{SM}}$ can be calculated, uses only the measurements of the UT side $R_b$ and the UT angle $\gamma$, as illustrated in Fig.\ \ref{fig:SM_Apex}.
The side $R_b$ is defined as
\begin{equation}\label{eq:Rb}
    R_b \equiv \left(1-\frac{\lambda^2}{2}\right)\frac{1}{\lambda}\left|\frac{V_{ub}}{V_{cb}}\right|
    = \sqrt{\bar\rho\,^2 + \bar\eta\,^2}\:,
\end{equation}
where $\lambda \equiv |V_{us}|$, and $|V_{ub}|$ and $|V_{cb}|$ can be measured in semileptonic kaon and $B$-meson decays, respectively.
The angle $\gamma$ is determined from $B\to DK$ and $B\to D\pi$ decays, where the latest average \cite{Amhis:2019ckw} reads
\begin{equation}\label{eq:gamma}
    \gamma = (71.1_{-5.3}^{+4.6})^{\circ}\:.
\end{equation}
Both $R_b$ and $\gamma$ can thus be completely determined from decays with only tree topologies, which are generally considered to be free from NP contributions, a hypothesis we will assume throughout this paper.
For the value of $\gamma$ in Eq.\ \eqref{eq:gamma}, Fig.\ \ref{fig:SM_Apex} shows that the precision of $\beta$, and thus also $\phi_d^{\text{SM}}$, is fully governed by the uncertainty on $R_b$.
Unfortunately, we also encounter difficulties with the determination of $R_b$ due to unresolved tensions between the various measurements, as extensively discussed in the literature and summarised in the reviews of Ref.\ \cite{Zyla:2020zbs}.
Here, we would like to reiterate some of the open issues, focusing on the impact they have on the SM predictions for $\phi_d$ and $\phi_s$.

Firstly, the CKM element $|V_{us}|$ is most precisely measured in semileptonic kaon decays.
The experimental average from $K\ell3$-type decays, with three particles in the final state, is given by \cite{Moulson:2017ive}
\begin{equation}
    |V_{us}|f_+(0) = 0.2165 \pm 0.0004\:,
\end{equation}
which in combination with the latest calculation of the form factor $f_+(0)$ from the Flavour Lattice Averaging Group (FLAG) \cite{Aoki:2019cca} gives
\begin{equation}\label{eq:Vus}
    |V_{us}| = 0.2231 \pm 0.0007\:.
\end{equation}
The experimental average \cite{Zyla:2020zbs} from $K\to\mu\nu_{\mu}\gamma$ decays ($K\ell2$-type) is
\begin{equation}
    |V_{us}| = 0.2252 \pm 0.0005\:,
\end{equation}
and differs from the result \eqref{eq:Vus} by three standard deviations.
Using the average of both results, even with inflated uncertainties to account for their discrepancy, leads to a three sigma deviation from unity \cite{Zyla:2020zbs} in an experimental test of the orthogonality relation
\begin{equation}
    |V_{ud}|^2 + |V_{us}|^2 + |V_{ub}|^2 = 1
\end{equation}
of the CKM matrix.

Secondly, for the CKM elements $|V_{ub}|$ and $|V_{cb}|$ there is a well-known discrepancy between the results obtained from inclusive and exclusive measurements (see Ref.\ \cite{Gambino:2020jvv} for a detailed discussion).
The latest averages from the Heavy Flavour Averaging Group (HFLAV) \cite{Amhis:2019ckw} for the exclusive determination of $|V_{ub}|$ and $|V_{cb}|$ are
\begin{equation}\label{eq:Vb_excl}
    |V_{ub}|_{\text{excl}} = (3.49 \pm 0.13) \times 10^{-3}\:,\qquad
    |V_{cb}|_{\text{excl}} = (39.25 \pm 0.56) \times 10^{-3}\:,
\end{equation}
which includes the constraint from $\Lambda_b^0\to p\mu^-\bar\nu_{\mu}$ decays \cite{Aaij:2015bfa}.
For the inclusive determination, on the other hand, the Gambino--Giordano--Ossola--Uraltsev (GGOU) \cite{Gambino:2007rp} approach for $|V_{ub}|$ and the $|V_{cb}|$ calculation using the kinematic scheme give \cite{Amhis:2019ckw}
\begin{equation}
    |V_{ub}|_{\text{incl}} = (4.32 \pm 0.17) \times 10^{-3}\:,\qquad
    |V_{cb}|_{\text{incl}} = (42.19 \pm 0.78) \times 10^{-3}\:.
\end{equation}
Combining the measurements of $|V_{us}|$, $|V_{ub}|$ and $|V_{cb}|$ results in four independent values for the UT side $R_b$
\begin{align}
    R_{b,\text{excl},K\ell3} & = 0.389 \pm 0.016\:, &
    R_{b,\text{incl},K\ell3} & = 0.448 \pm 0.019\:, \label{eq:Rb_excl} \\
    R_{b,\text{excl},K\ell2} & = 0.385 \pm 0.015\:, &
    R_{b,\text{incl},K\ell2} & = 0.443 \pm 0.019\:.
\end{align}
and a difference between the inclusive and exclusive determinations at the level of two standard deviations.

\begin{figure}
    \centering
    \includegraphics[width=0.49\textwidth]{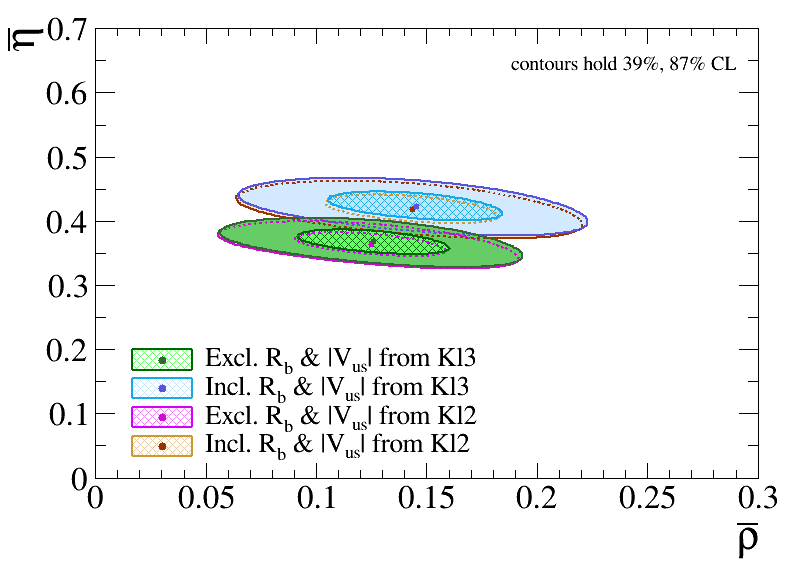}
    \includegraphics[width=0.49\textwidth]{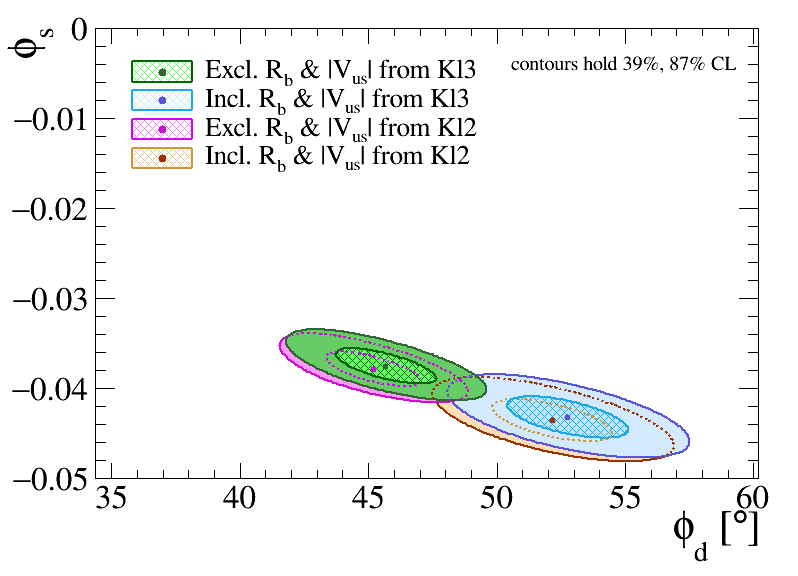}
    \caption{Two-dimensional confidence regions of the four SM predictions for the UT apex $(\bar\rho,\bar\eta)$ (left) and for $\phi_d$ and $\phi_s$ (right), based on different choices for $\lambda$ and $R_b$.}
    \label{fig:phi_SM}
\end{figure}

A fit to the measurements of $\lambda$, $R_b$ and $\gamma$ is performed to determine the UT apex $(\bar\rho,\bar\eta)$, or directly the mixing phases $\phi_d$ and $\phi_s$, with the results shown in Fig.\ \ref{fig:phi_SM}.
This fit is implemented using the GammaCombo framework \cite{Aaij:2016kjh}, originally developed by the LHCb collaboration as a statistical framework to combine their various measurements of the UT angle $\gamma$.
From Fig.\ \ref{fig:phi_SM}, it becomes clear that the discrepancy between the inclusive and exclusive determinations of $|V_{ub}|$ and $|V_{cb}|$ is the dominant source of uncertainty for both the apex $(\bar\rho,\bar\eta)$ and the mixing phases $\phi_d$ and $\phi_s$.
But more surprisingly, also the choice of $\lambda$ has a non-negligible impact on the SM predictions.
The numerical results for the mixing phases are
\begin{align}
    \text{Excl,} K\ell3 & &
    \phi_d^{\text{SM}} & = (45.7 \pm 2.0)^{\circ}\:, & 
    \phi_s^{\text{SM}} & = -0.0376 \pm 0.0020 = (-2.15 \pm 0.11)^{\circ}\:,\label{eq:phi_SM_E3} \\
    \text{Excl,} K\ell2 & &
    \phi_d^{\text{SM}} & = (45.2 \pm 1.8)^{\circ}\:, & 
    \phi_s^{\text{SM}} & = -0.0379 \pm 0.0020 = (-2.18 \pm 0.11)^{\circ}\:,\\
    \text{Incl,} K\ell3 & &
    \phi_d^{\text{SM}} & = (52.7 \pm 2.4)^{\circ}\:, & 
    \phi_s^{\text{SM}} & = -0.0433 \pm 0.0024 = (-2.49 \pm 0.14)^{\circ}\:,\\
    \text{Incl,} K\ell2 & &
    \phi_d^{\text{SM}} & = (52.1 \pm 2.4)^{\circ}\:, & 
    \phi_s^{\text{SM}} & = -0.0436 \pm 0.0024 = (-2.52 \pm 0.14)^{\circ}\:,\label{eq:phi_SM_I2}
\end{align}
For $\phi_s^{\text{SM}}$, this result is a factor 2.5 to 3 less precise than the value
\cite{Charles:2015gya}
\begin{equation}
    \phi_s^{\text{SM}} = -0.03696_{-0.00072}^{+0.00084} = \left(-2.118_{-0.041}^{+0.048}\right)^{\circ}\:,
\end{equation}
obtained from the global fit of the UT, typically used in the literature.

%
%
%
\section{Theoretical Framework}\label{sec:PenFrame}
%
%
%
\subsection{Decay Amplitudes}
The transition amplitudes of the five \BJpsiX decays discussed in this paper are dominated by the contribution from the colour-suppressed tree topology, parametrised by a CP-conserving amplitude ``$C$''.
They also receive contributions from penguin topologies ``$P^{(q)}$'', where $q=u,c,t$ labels the exchanged quark flavour, and in the case of the \mbox{\BdJpsiPi,} \BdJpsiRho and \BsJpsiPhi decays from exchange and penguin-annihilation diagrams.
The latter two are expected to be even smaller than the penguin topologies, and will therefore be neglected in the present analysis.
The \BdJpsiPhi channel only gets contributions from the exchange and penguin-annihilation topologies, and its branching fraction can thus be used to probe these diagrams.
LHCb has recently put an upper limit on the branching fraction of this decay of $1.1\times 10^{-7}$ at 90\% confidence level \cite{Aaij:2020dlx}, which supports the assumed hierarchy between the decay topologies and our choice to neglect exchange and penguin-annihilation contributions.

We will also neglect potential NP contributions to the decay amplitudes, thus only allowing NP to enter via the $B_q^0$--$\bar B_q^0$ mixing phase $\phi_q$.
In this way, we can make use of the SM structure, and in particular the unitarity of the CKM matrix, to express the transition amplitudes for the decay of a neutral $B$ meson into a CP eigenstate $f$ in the following form \cite{Fleischer:1999zi}:
\begin{align}
    A(B_q^0\to f) & \equiv \phantom{\eta_f}\mathcal{N}_f\left[1-b_f e^{i\rho_f}e^{+i\gamma}\right]\:, \label{eq:TransAmp} \\
    A(\bar B_q^0\to f) & \equiv \eta_f\mathcal{N}_f\left[1-b_f e^{i\rho_f}e^{-i\gamma}\right]\:,
\end{align}
where $\eta_f$ is the CP-eigenvalue of the final state $f$.
In these expressions, $\mathcal{N}_f$ is a CP-conserving normalisation factor which is governed by the dominant tree topology, while $b_f$ gives the relative contribution of the penguin topologies with respect to the tree contribution.
The CP-conserving strong phase difference between both terms is parametrised as $\rho_f$, whereas the relative weak phase is given by the UT angle $\gamma$.

For the \BdJpsiKS (or \BdJpsiKL) decay we have to substitute 
\begin{equation}\label{eq:sub_ccs}
    \mathcal{N}_f \to \left(1-\frac{\lambda^2}{2}\right)\mathcal{A}'\:, \qquad
    b_f e^{i\rho_f} \to -\epsilon a' e^{i\theta'}
\end{equation}
in the transition amplitude \eqref{eq:TransAmp}, which then takes the following form \cite{Fleischer:1999nz}:
\begin{equation}\label{eq:TransAmp_Bd2JpsiK}
    A\left(\BdJpsiKS\right) = \left(1-\frac{\lambda^2}{2}\right)\mathcal{A}' \left[1+\epsilon a'e^{i\theta'}e^{i\gamma}\right]\:.
\end{equation}
The primes $(\prime)$ are introduced to distinguish these $\bar b\to \bar s c \bar c$ quark-level processes (primed) from their $\bar b\to d c\bar c$ counterparts (unprimed) discussed below.
The CKM factor $\epsilon$ gives rise to the suppression of the penguin effects in the $\bar b\to \bar s c \bar c$ transitions.
It can be expressed in terms of the Wolfenstein parameter $\lambda$ as
\begin{equation}
    \epsilon \equiv \frac{\lambda^2}{1-\lambda^2} = 0.05238 \pm 0.00035\:,
\end{equation}
where the numerical value is based on the measurement \eqref{eq:Vus}.
The hadronic amplitude $\mathcal{A}'$ and the penguin parameter $a'e^{i\theta'}$ can be decomposed in terms of the hadronic matrix elements associated with the tree and penguin topologies as
\begin{equation}\label{eq:HadAmp_ccs}
    \mathcal{A}' \equiv \lambda^2 A \left[C'+P^{\prime(c)}-P^{\prime(t)}\right]
\end{equation}
and 
\begin{equation}\label{eq:penguin_ccs}
    a'e^{i\theta'} \equiv R_b\left[\frac{P^{\prime(u)}-P^{\prime(t)}}{C'+P^{\prime(c)}-P^{\prime(t)}}\right]\:,
\end{equation}
where $R_b$ is defined in Eq.\ \eqref{eq:Rb}, and
\begin{equation}
    A \equiv \frac{|V_{cb}|}{\lambda^2}\:,
\end{equation}
are combinations of the relevant CKM matrix elements.
The UT side $R_b$ provides a natural scale for the size $a'$ of the penguin contributions: in a hypothetical scenario without loop suppression, the penguin topologies could be of similar size as the tree topology, i.e.\ $P'\approx C'$, and we would get $a'\approx R_b$.

The \BsJpsiPhi decay has two vector mesons in the final state, resulting in more complicated decay dynamics where the hadronic parameters depend on the final-state configuration.
This system can be described with three polarisation states: The CP-even eigenstates 0 and $\parallel$, and the CP-odd eigenstate $\perp$.
The three states can be disentangled through the angular distribution of the decay products of the vector mesons.
For each of these final states, the transition amplitude has a structure that is equivalent to the expression in Eq.~\eqref{eq:TransAmp_Bd2JpsiK}, where the hadronic amplitude $\mathcal{A}'_f$ and the penguin parameters $a'_f$, $\theta_f'$ should in principle be considered for each polarisation state $f$ individually.
Applying naive factorisation for the hadronic matrix elements of the four-quark operators, the penguin parameters do not depend on the final-state configuration $f$ \cite{Fleischer:1999zi}.
Since experimental analyses of CP violation in these decays have so far focused on polarisation-independent measurements, we will do the same in the analysis of the current data.
However, we hope that future updates of these measurements will make a polarisation-dependent analysis possible.
In addition, it is important to distinguish the strong interaction effects in the vector--pseudo-scalar and vector--vector decays \BdJpsiK and \BsJpsiPhi as they have different decay dynamics.
We will label the penguin parameters arising in the latter channel as $a'_V$ and $\theta'_V$ (suppressing a dependence on the final-state configuration of the vector mesons).

The transition amplitude for the \BsJpsiKS decay is obtained by substituting
\begin{equation}\label{eq:sub_ccd}
    \mathcal{N}_f \to -\lambda \mathcal{A}\:, \qquad
    b_f e^{i\rho_f} \to a e^{i\theta}\:,
\end{equation}
leading to \cite{Fleischer:1999nz}:
\begin{equation}\label{eq:TransAmp_BsJpsiK}
    A\left(\BsJpsiKS\right) = - \lambda \mathcal{A} \left[1- a e^{i\theta}e^{i\gamma}\right]\:,
\end{equation}
where the hadronic parameters are defined in analogy to Eqs.~\eqref{eq:HadAmp_ccs} and \eqref{eq:penguin_ccs}.
It should be noted that -- in contrast to Eq.~\eqref{eq:TransAmp_Bd2JpsiK} -- there is no factor $\epsilon$ in front of the second term, thereby amplifying the penguin effects in the $\bar b\to \bar d c \bar c$ modes with respect to their $\bar b\to \bar s c \bar c$ counterparts.
However, the overall amplitude is suppressed by a factor $\lambda$, which reduces the decay rate and makes these decays experimentally more challenging to study.
For the \BdJpsiPi and \BdJpsiRho modes, the transition amplitude has a structure that is equivalent to the expression in Eq.~\eqref{eq:TransAmp_BsJpsiK}.
In the case of the \BdJpsiRho channel, an angular analysis of the decay products of the vector mesons is needed, similar to the \BsJpsiPhi decay \cite{Fleischer:1999zi}.

The $SU(3)$ flavour symmetry of the strong interaction allows us to relate the hadronic parameters of the $\bar b\to \bar s c \bar c$ and $\bar b\to \bar d c \bar c$ transitions to one another, yielding
\begin{equation}\label{eq:SU3_pen}
    a'e^{i\theta'} = ae^{i\theta}
\end{equation}
as well as
\begin{equation}\label{eq:SU3_had}
    \mathcal{A}'=\mathcal{A}\:.
\end{equation}
But because $m_u \approx m_d < m_s$, the $SU(3)$ flavour symmetry does not hold perfectly, and the relations \eqref{eq:SU3_pen} and \eqref{eq:SU3_had} get $SU(3)$-breaking corrections.
In the factorisation approximation, the hadronic form factors and decay constants cancel in the ratio \eqref{eq:penguin_ccs}.
Consequently, the $SU(3)$-breaking corrections can only enter relation \eqref{eq:SU3_pen} through non-factorisable effects.
Such a cancellation does not happen for the hadronic amplitudes \eqref{eq:HadAmp_ccs}, and the relation \eqref{eq:SU3_had} can thus get both factorisable and non-factorisable corrections.

Information on the penguin parameters is encoded in the CP asymmetries as well as the branching fraction of the decay.
The former depend only on the parameters $b_f$ and $\rho_f$, while the latter also involves the normalisation factor $\mathcal{N}_f$.
Although it is possible to calculate this hadronic amplitude within the factorisation approximation and thus use the branching fraction measurements to help constrain the penguin parameters (see Ref.~\cite{DeBruyn:2014oga} for an example), this approach suffers from the corresponding theoretical uncertainties.
To avoid this limitation and determine $\phi_d$ and $\phi_s$ with the highest possible precision, we will only use the CP asymmetries, which are unaffected by theoretical uncertainties due to factorisation, to determine the penguin parameters.
We shall return to the discussion of the branching fraction information in Section \ref{sec:BR}, utilising it to obtain insights into the hadron dynamics of the relevant decays.

\subsection{CP Asymmetries}

The time-dependent CP asymmetry for neutral $B_q$ mesons is given by
\begin{align}
    a_{\text{CP}}(t) & \equiv
    \frac{|A(B_q^0(t)\to f)|^2-|A(\bar B_q^0(t)\to f)|^2}{|A(B_q^0(t)\to f)|^2+|A(\bar B_q^0(t)\to f)|^2} \\
    & = \frac{\mathcal{A}_{\text{CP}}^{\text{dir}}\cos(\Delta m_qt)+\mathcal{A}_{\text{CP}}^{\text{mix}}\sin(\Delta m_qt)}{\cosh(\Delta\Gamma_qt/2)+\mathcal{A}_{\Delta\Gamma}\sinh(\Delta\Gamma_qt/2)}\:,
\end{align} 
where $\Delta m_q\equiv m^{(q)}_{\text{H}}-m^{(q)}_{\text{L}}$ and $\Delta\Gamma_q\equiv \Gamma_{\text{L}}^{(q)}-\Gamma_{\text{H}}^{(q)}$ are the mass and decay width difference between the heavy and light eigenstates of the $B_q$-meson system, respectively.
The direct ($\mathcal{A}_{\text{CP}}^{\text{dir}}$) and mixing-induced ($\mathcal{A}_{\text{CP}}^{\text{mix}}$) CP asymmetries depend on the penguin parameters $b_f$ and $\rho_f$, and the $B_q^0$--$\bar B_q^0$ mixing phase $\phi_q$ as follows \cite{Fleischer:1999zi}:
\begin{align}
    \mathcal{A}_{\text{CP}}^{\text{dir}}(B_q\to f) & = \frac{2 b_f \sin\rho_f\sin\gamma}{1-2b_f\cos\rho_f\cos\gamma+b_f^2}\:, \label{eq:Adir}\\
    \eta_f\mathcal{A}_{\text{CP}}^{\text{mix}}(B_q\to f) & = \left[ \frac{\sin\phi_q-2 b_f \cos\rho_f\sin(\phi_q+\gamma)+b_f^2\sin(\phi_q+2\gamma)}{1 - 2b_f\cos\rho_f\cos\gamma+b_f^2}\right]\:.\label{eq:Amix}
\end{align}
These observables can thus be used to determine the three parameters of interest.
In the discussion below we have chosen to always reference the quantity $\eta\mathcal{A}_{\text{CP}}^{\text{mix}}$ as it is independent of the CP-eigenvalue of the final state and therefore easier to combine with other measurements.
The label $f$ identifying the final state has been dropped from $\eta_f$ to simplify the notation.
The mass eigenstate rate asymmetry
\begin{equation}
    \eta_f\mathcal{A}_{\Delta\Gamma}(B_q\to f) = -\left[\frac{\cos\phi_q-2 b_f \cos\rho_f\cos(\phi_q+\gamma)+b_f^2\cos(\phi_q+2\gamma)}{1 - 2b_f\cos\rho_f\cos\gamma+b_f^2}\right]
\end{equation}
depends also on the penguin parameters, but it is not independent from the direct and mixing-induced CP asymmetries, satisfying the relation
\begin{equation}
    \left[\mathcal{A}_{\text{CP}}^{\text{dir}}(B_q\to f)\right]^2 +
    \left[\mathcal{A}_{\text{CP}}^{\text{mix}}(B_q\to f)\right]^2  +
    \left[\mathcal{A}_{\Delta\Gamma}(B_q\to f)\right]^2 = 1\:.
\end{equation}

In the absence of the doubly Cabibbo-suppressed penguin contributions, i.e.\ $b_f = 0$, these expressions simplify to the familiar forms 
\begin{equation}
    \mathcal{A}_{\text{CP}}^{\text{dir}} = 0\:,\qquad
    \eta_f\mathcal{A}_{\text{CP}}^{\text{mix}} = \sin\phi_q\:, 
\end{equation}
which would allow us to determine $\phi_q$ directly from the mixing-induced CP asymmetry.
On the other hand, allowing for the penguin effects, i.e.\ $b_f \neq 0$, the CP asymmetries are related to the effective mixing phase introduced in Eq.~\eqref{eq:eff_mix_phase}, with the complete relation given as follows \cite{Faller:2008gt}:
\begin{equation}\label{eq:DeltaPhi}
    \frac{\eta_f \mathcal{A}_{\text{CP}}^{\text{mix}}(B_q\to f)}{\sqrt{1 - \left(\mathcal{A}_{\text{CP}}^{\text{dir}}(B_q\to f)\right)^2}}
    =\sin(\phi_q+\Delta\phi_q^f) \equiv \sin(\phi_{q,f}^{\text{eff}})\:.
\end{equation}
The phase shift $\Delta\phi_q^f$ is defined in terms of the penguin parameters $b_f$ and $\rho_f$ as
\begin{align}
    \sin \Delta\phi_q^f & =\frac{-2b_f\cos\rho_f\sin\gamma+b_f^2\sin2\gamma}{\left(1-2b_f\cos\rho_f\cos\gamma+b_f^2\right)\sqrt{1- \left(\mathcal{A}_{\text{CP}}^{\text{dir}}(B\to f)\right)^2}}\:,\\
    \cos \Delta\phi_q^f & =\frac{1-2b_f\cos\rho_f\cos\gamma+b_f^2\cos2\gamma}{\left(1-2b_f\cos\rho_f\cos\gamma+b_f^2\right)\sqrt{1- \left(\mathcal{A}_{\text{CP}}^{\text{dir}}(B\to f)\right)^2}}\:,
\end{align}
yielding
\begin{equation}
    \tan \Delta\phi_q^f = -\left[ \frac{2b_f\cos\rho_f\sin\gamma-b^2\sin2\gamma}{1-2b_f\cos\rho_f\cos\gamma+b_f^2\cos2\gamma} \right]\:.
\end{equation}
Additional information about the penguin contributions is thus necessary to correctly interpret the experimental measurements and determine the mixing phase $\phi_q$.
It is important to distinguish these phases from the effective ones governing the mixing-induced CP asymmetries.
If we knew the hadronic penguin parameters, we could straightforwardly calculate the hadronic phase shifts $\Delta\phi_q^f$ with the expressions given above.
This correction is often ignored in the literature.

%
%
%
\section{Picture from Current Data}\label{sec:CurrData}
Let us now explore the picture emerging from the current data, and extract the values of the CP-violating phases $\phi_q$, which is a key focus of this paper.

\subsection[Determination of phid]{Determination of $\phi_d$}\label{sec:Data_phi_d}

The $B_d^0$--$\bar B_d^0$ mixing phase $\phi_d$ is determined from the \BdJpsiKS mixing-induced CP asymmetry.
The penguin parameters $a$ and $\theta$, which are needed to take the shift $\Delta\phi_d$ in Eq.~\eqref{eq:DeltaPhi} into account, can be determined in a theoretically clean way through the $U$-spin partner \BsJpsiKS \cite{DeBruyn:2014oga}.
Although this is \emph{the} preferred strategy to obtain the highest precision of $\phi_d$ in the upgrade era of the LHCb and Belle II experiments, the current experimental uncertainties on the CP asymmetries \cite{Aaij:2015tza}:
\begin{equation}\label{eq:BsJpsiK}
    \mathcal{A}_{\text{CP}}^{\text{dir}}(\BsJpsiKS) = -0.28 \pm 0.42\:,\qquad
    \eta\mathcal{A}_{\text{CP}}^{\text{mix}}(\BsJpsiKS) = 0.08 \pm 0.41\:,
\end{equation}
are unfortunately still too large to constrain $a$ and $\theta$ in a meaningful way.
However, stronger constraints on the penguin effects can already be obtained using the data for the \mbox{\BdJpsiPi} decay \cite{Faller:2008zc}.

\begin{figure}
    \centering
    \includegraphics[width=0.49\textwidth]{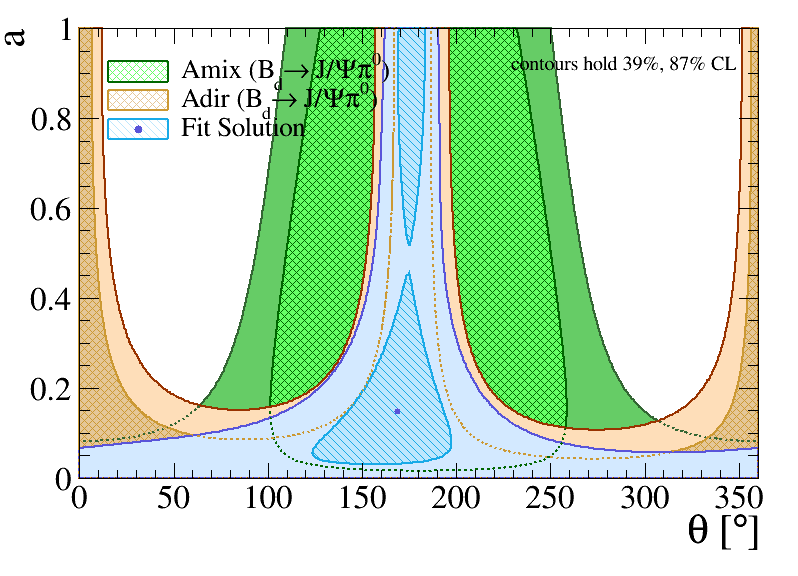}
    \includegraphics[width=0.49\textwidth]{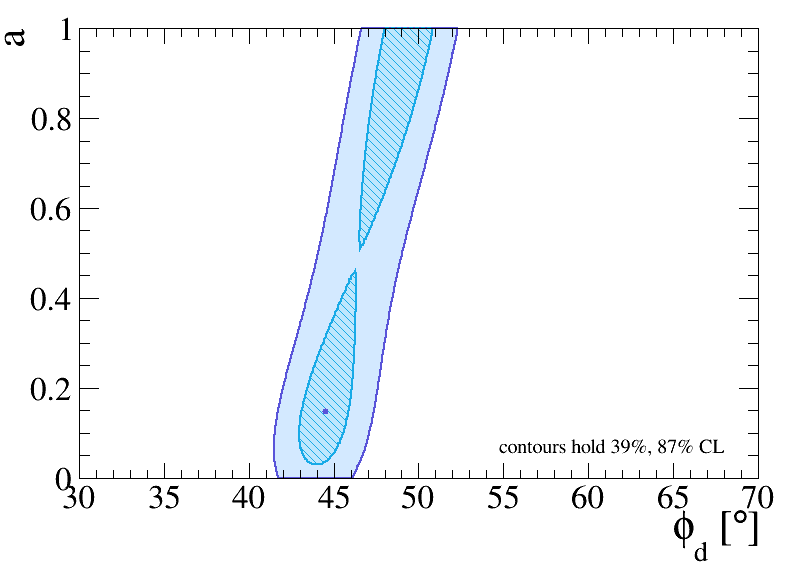}
    \caption{Two-dimensional confidence regions of the fit for the penguin parameters and $\phi_d$ from the CP asymmetries in \BdJpsiPi and \BdJpsiK.
    Note that the contours for $\mathcal{A}_{\text{CP}}^{\text{dir}}(\BdJpsiPi)$ and $\mathcal{A}_{\text{CP}}^{\text{mix}}(\BdJpsiPi)$ are added for illustration only.
    They include the best fit solutions for $\phi_d$ and $\gamma$ as Gaussian constraints.
    }
    \label{fig:Bd2JpsiX}
\end{figure}

Using the latest experimental averages for $\gamma$ and $\phi_d$ as external constraints, the penguin parameters can be determined from the CP asymmetries of the \BdJpsiPi channel, which are given by the following results from HFLAV \cite{Amhis:2019ckw}:
\begin{equation}\label{eq:Bd2JpsiPi}
    \mathcal{A}_{\text{CP}}^{\text{dir}}(\BdJpsiPi) =  0.04 \pm 0.12\:,\qquad
    \eta\mathcal{A}_{\text{CP}}^{\text{mix}}(\BdJpsiPi) = 0.86 \pm 0.14\:.
\end{equation}
However, the external input on $\phi_d$ would need to be corrected for potential penguin effects, which we aim to quantify here using \BdJpsiPi.
This strategy thus necessarily requires an iterative approach.
Instead, and because the experimental average of $\phi_d$ is dominated by the input from \BdJpsiKS, we perform a combined fit to the CP asymmetries of the \BdJpsiPi and \BdJpsiK channels to determine $a$, $\theta$ and the penguin-corrected value of $\phi_d$ simultaneously.
Neglecting differences due to CP violation in the neutral kaon system, which can in principle be accounted for, the decay modes \BdJpsiKS and \BdJpsiKL have the same decay structure and can thus straightforwardly be combined with each other.
These two channels only differ in the CP-eigenvalue of the final states, which is accounted for in the observable $\eta\mathcal{A}_{\text{CP}}^{\text{mix}}$.
The experimental averages \cite{Amhis:2019ckw} for the \BdJpsiK CP asymmetries used in this analysis are
\begin{equation}\label{eq:Bd2JpsiK}
    \mathcal{A}_{\text{CP}}^{\text{dir}}(\BdJpsiK) =  -0.007 \pm 0.018\:,\qquad
    \eta\mathcal{A}_{\text{CP}}^{\text{mix}}(\BdJpsiK) = 0.690 \pm 0.018\:,
\end{equation}
and correspond to an effective mixing phase
\begin{equation}\label{eq:phi_d_eff}
    \phi_{d,J/\psi K^0}^{\text{eff}} = (43.6 \pm 1.4)^{\circ}\:.
\end{equation}
In principle, the four inputs in Eqs.\ \eqref{eq:Bd2JpsiPi} and \eqref{eq:Bd2JpsiK} provide sufficient information to also determine the UT angle $\gamma$, but the corresponding precision is not competitive with other direct measurements \cite{DeBruyn:2010hh}.
It is therefore more advantageous to still add the average \eqref{eq:gamma} as an external constraint.
In order to relate the penguin parameters in \BdJpsiPi and \BdJpsiK with one another, the fit assumes the $SU(3)$ relation \eqref{eq:SU3_pen}, neglects contributions from exchange and penguin-annihilation topologies as well as non-factorisable $SU(3)$-breaking effects.
From the fit, implemented in the GammaCombo framework \cite{Aaij:2016kjh}, we obtain
\begin{equation}\label{eq:Bd2JpsiX}
    a = 0.15_{-0.12}^{+0.31}\:, \qquad
    \theta = \left(168_{-47}^{+31}\right)^{\circ}\:, \qquad
    \phi_d = \left(44.5_{-1.5}^{+1.8}\right)^{\circ}\:.
\end{equation}
Due to the non-trivial dependence of the CP asymmetries on the penguin parameters, these uncertainties are highly non-Gaussian, as also illustrated by the two-dimensional confidence regions in Fig.~\ref{fig:Bd2JpsiX}.
This is true for all results presented in Sections \ref{sec:CurrData}, \ref{sec:BR} and \ref{sec:bench} derived from the fits of the penguin parameters.

In comparison with $\phi_{d,J/\psi K^0}^{\text{eff}}$ in \eqref{eq:phi_d_eff}, the uncertainty of $\phi_d$ is slightly larger due to its correlation with the penguin parameters, as illustrated by the two-dimensional confidence regions in Fig.~\ref{fig:Bd2JpsiX}.
The solution for $a$ and $\theta$ corresponds to the phase shift
\begin{equation}
    \Delta\phi_d = \left(-0.8_{-1.8}^{+0.7}\right)^{\circ}\:.
\end{equation}
The two-dimensional confidence regions given in Fig.~\ref{fig:Bd2JpsiX} show a second solution with $a\approx 1$.
However, looking at the definition of the penguin parameter $a$ in Eq.~\eqref{eq:penguin_ccs}, we observe that a solution with $a$ larger than the UT side $R_b$ would correspond to penguin contributions much larger than the tree amplitude, which is highly disfavoured.
The presence of this second solution is a direct consequence of the absence of direct CP violation in the \BdJpsiPi and \BdJpsiK channels, which leads to a preferred solution for the phase $\theta$ around $180^{\circ}$.
This in turn limits the sensitivity of the current data to constrain the size $a$ of the penguin effects.
Instead of using the arguments above, the two-fold ambiguity can also be resolved by including the CP asymmetries of the \BsJpsiKS channel in the fit, as will be shown in the combined fit for $\phi_d$ and $\phi_s$ below.

The plot on the right-hand side in Fig.~\ref{fig:Bd2JpsiX} shows a strong correlation between $a$ and the CP-violating phase $\phi_d$, which highlights the importance of controlling the penguin effects in order to obtain the highest precision of $\phi_d$, both from an experimental and from a theoretical point of view.

With the current experimental precision, the two-dimensional constraints in the $\theta$--$a$ plane coming from the direct CP asymmetries of \BdJpsiPi and \BdJpsiK completely overlap.
Consequentially, our analysis is not sensitive to possible $SU(3)$-breaking effects between $a'e^{i\theta'}$ and $ae^{i\theta}$.
A combined analysis of the \BdJpsiPi and \BdJpsiK CP asymmetries will only reveal $SU(3)$ breaking between both decay channels when the two-dimensional constraints from the direct CP asymmetries are incompatible.
Experimentally establishing a non-zero direct CP asymmetry in \BdJpsiPi is a necessary -- but not sufficient -- condition for this to happen.
For the central value in Eq.~\eqref{eq:Bd2JpsiPi}, this requires an order of magnitude improvement in the experimental precision.
The impact of $SU(3)$ breaking can therefore safely be ignored in the present analysis, but should be re-evaluated in future updates.

\subsection[Determination of phis]{Determination of $\phi_s$}

\begin{figure}
    \centering
    \includegraphics[width=0.49\textwidth]{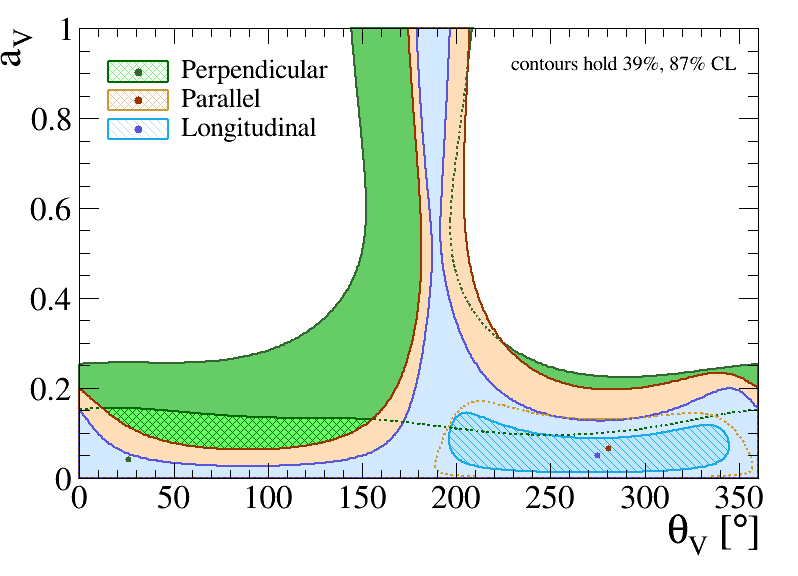}
    \caption{Comparison of the two-dimensional confidence regions of the fit for the penguin parameters from the polarisation-dependent CP asymmetries in \BdJpsiRho.}
    \label{fig:Bd2JpsiRho}
\end{figure}

The counterpart of the golden mode \BdJpsiKS for the $B_s^0$--$\bar{B}_s^0$ mixing phase $\phi_s$ is the decay \BsJpsiPhi, which is related through the exchange of the spectator down quark with a strange quark.
In contrast to the former channel, the latter has two vector mesons in the final state and its decay is hence described by three polarisation states (0, $\parallel$, $\perp$), as mentioned earlier.
In the most ideal scenario for the theoretical interpretation of the data, we would have individual measurements of the direct and mixing-induced CP asymmetries for all three polarisation states, as this would allow us to correct for polarisation-dependent hadronic effects.
However, this makes the fit to the data much more challenging, and the experiments have so far opted to report a single effective mixing phase $\phi_{s,J/\psi\phi}^{\text{eff}}$ instead.
Some analyses have explored polarisation-dependent results for $\phi_{s,J/\psi\phi}^{\text{eff}}$ \cite{Aaij:2014zsa, Aaij:2019vot}, but this has not yet become the baseline.
A second experimental challenge are the contributions from the $f_0(980)$ and other light resonances in the \mbox{$B_s^0\to J/\psi K^+K^-$} final state, which have been studied in detail in Ref.\ \cite{Aaij:2013orb}.
They can be disentangled through an angular analysis of the final state particles, and the state-of-the-art experimental analyses now include a background $S$-wave component to account for them.

Averaging the measurements from D0 \cite{Abazov:2011ry}, CDF \cite{Aaltonen:2012ie}, CMS \cite{Khachatryan:2015nza} and ATLAS \cite{Aad:2020jfw}, which all assume
\begin{equation}
    |\lambda| \equiv \left| \frac{A(B_q^0\to f)}{A(\bar B_q^0\to f)} \right| = 1\:,
\end{equation}
with the measurement from LHCb \cite{Aaij:2019vot}, which also measured $|\lambda| = 0.994 \pm 0.013$, we get
\begin{equation}
    \phi_{s,J/\psi\phi}^{\text{eff}} = -0.085 \pm 0.025 = (-4.9 \pm 1.4)^{\circ}\:.
\end{equation}
However, for the analysis of the penguin effects it is more convenient to convert the LHCb measurements of $|\lambda|$ and the experimental average for $\phi_{s,J/\psi\phi}^{\text{eff}}$ back into the CP asymmetries:
\begin{equation}\label{eq:BsJpsiPhi}
    \mathcal{A}_{\text{CP}}^{\text{dir}}(\BsJpsiPhi) = 0.006 \pm 0.013\:,\qquad
    \mathcal{A}_{\text{CP}}^{\text{mix}}(\BsJpsiPhi) = -0.085 \pm 0.025\:.
\end{equation}
Note that because $|\lambda|$ is compatible with unity, we get $\mathcal{A}_{\text{CP}}^{\text{mix}} = \sin(\phi_{s,J/\psi\phi}^{\text{eff}})$ to a very good approximation.

We assume that the $\phi$ meson is a pure $s\bar s$ state, and hence a superposition of an $SU(3)$ octet state and $SU(3)$ singlet state (see Ref.\ \cite{Faller:2008gt} for a detailed discussion).
Ignoring any contributions associated with the singlet state, the penguin effects in \BsJpsiPhi can be determined using the \BdJpsiRho decay as the control mode, as was previously discussed in Ref.~\cite{Fleischer:1999zi,DeBruyn:2014oga}.
The CP-violating observables of the \BdJpsiRho channel have been measured for all three polarisation states \cite{Aaij:2014vda}, allowing us to compare the confidence regions for the penguin parameters $a_V$ and $\theta_V$ between the longitudinal, parallel and perpendicular polarisation states, as shown in Fig.~\ref{fig:Bd2JpsiRho}.
Within the current precision, we find agreement between the three polarisation states and could not resolve differences, thereby setting the stage to continue with the determination of the penguin parameters affecting the polarisation-independent results for \BsJpsiPhi.
But it should be stressed again that improved precision of the input measurements, as can be expected from the upgrade programmes of LHCb and Belle II, may lead to observable differences in a comparison like Fig.~\ref{fig:Bd2JpsiRho}, thus also affecting the determination of $\phi_s$ from \BsJpsiPhi.
We will illustrate such a scenario in Section~\ref{sec:bench}.

\subsection[Simultaneous Analysis of phid and phis]{Simultaneous Analysis of $\phi_d$ and $\phi_s$}\label{sec:BJpsiX_fit}

\begin{figure}
    \centering
    \includegraphics[width=0.6\textwidth]{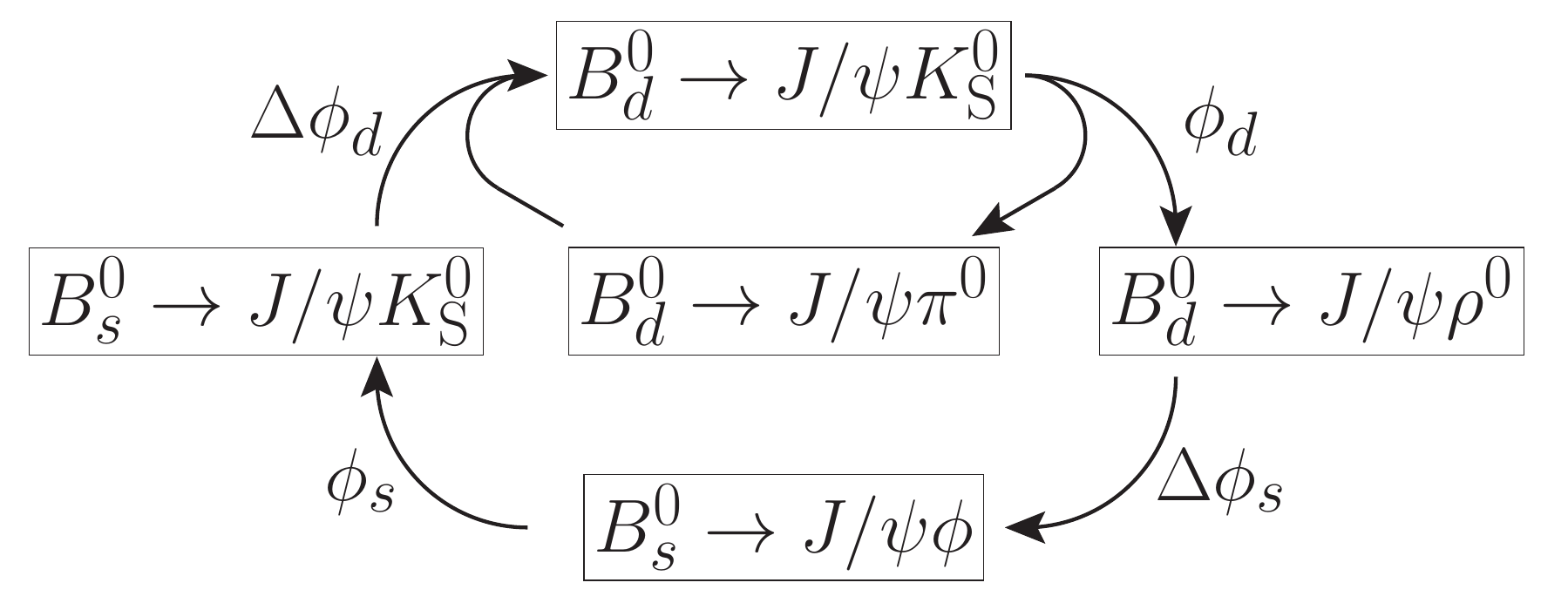}
    \caption{The cross-dependence between the determination of $\phi_d$ and $\phi_s$ and their hadronic penguin shifts, showing the interplay between the five \BJpsiX decays discussed in this paper.}
    \label{fig:Interplay}
\end{figure}

In analogy to the \BdJpsiPi analysis, the polarisation-independent CP asymmetries of the \BdJpsiRho channel, which take the following experimental values \cite{Aaij:2014vda}:
\begin{equation}\label{eq:Bd2JpsiRho}
    \mathcal{A}_{\text{CP}}^{\text{dir}}(\BdJpsiRho) =  -0.064 \pm 0.059\:,\qquad
    \eta\mathcal{A}_{\text{CP}}^{\text{mix}}(\BdJpsiRho) = 0.66 \pm 0.15\:,
\end{equation}
have to be complemented with external constraints for $\gamma$ and $\phi_d$ in order to determine the penguin parameters $a_V$ and $\theta_V$.
We could now use the result \eqref{eq:Bd2JpsiX} obtained above, which shows the cross-dependence of $\phi_d$ and $\phi_s$ on each another.
When using \BsJpsiKS to determine the penguin effects in the CP asymmetries of \BdJpsiK, this situation becomes circular, as illustrated in Fig.~\ref{fig:Interplay}: $\phi_s$ is required to determine the penguin shift $\Delta\phi_d$ from \BsJpsiKS, which is needed to extract $\phi_d$ from \BdJpsiK.
In turn $\phi_d$ is a necessary input to determine the penguin shift $\Delta\phi_s$ from \BdJpsiRho, which is needed to determine $\phi_s$ from the CP-violating asymmetries of the \BsJpsiPhi mode.
It should be emphasised again that $\phi_d$ and $\phi_s$ are the mixing phases themselves and not the effective ones which are affected by the penguin corrections.

\begin{figure}
    \centering
    \includegraphics[width=0.49\textwidth]{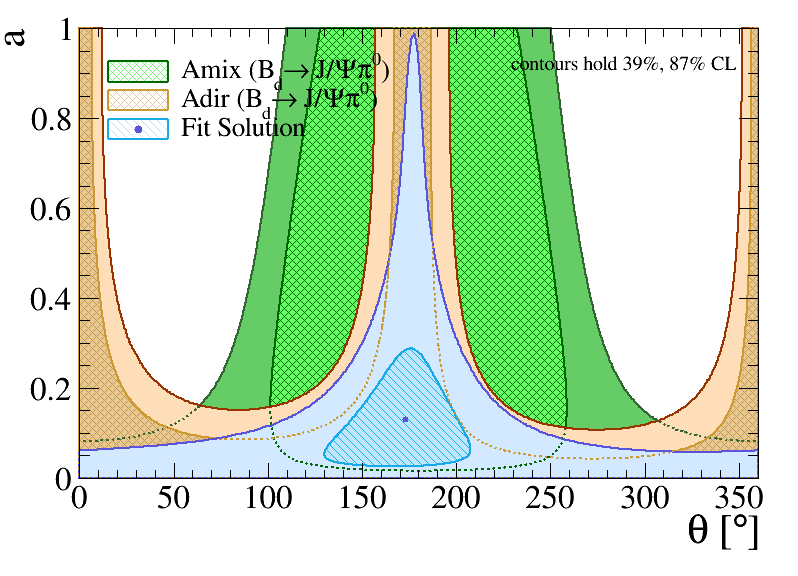}
    \includegraphics[width=0.49\textwidth]{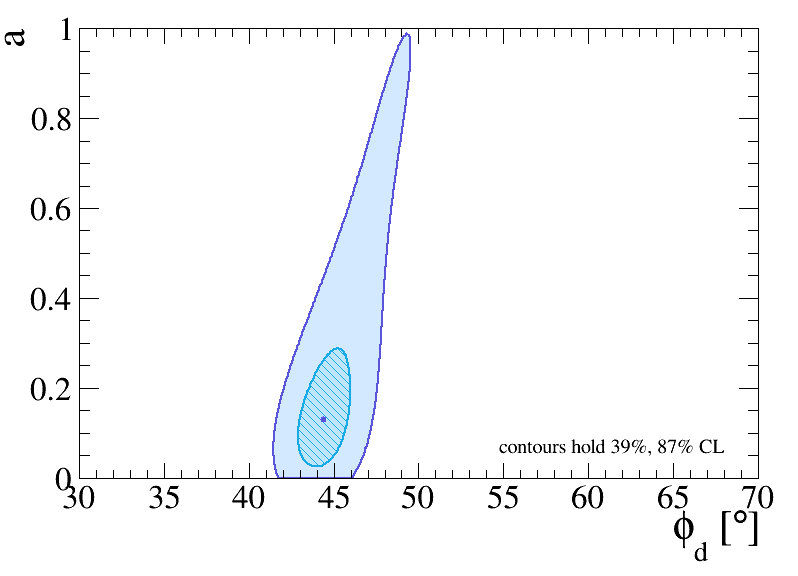}
    
    \includegraphics[width=0.49\textwidth]{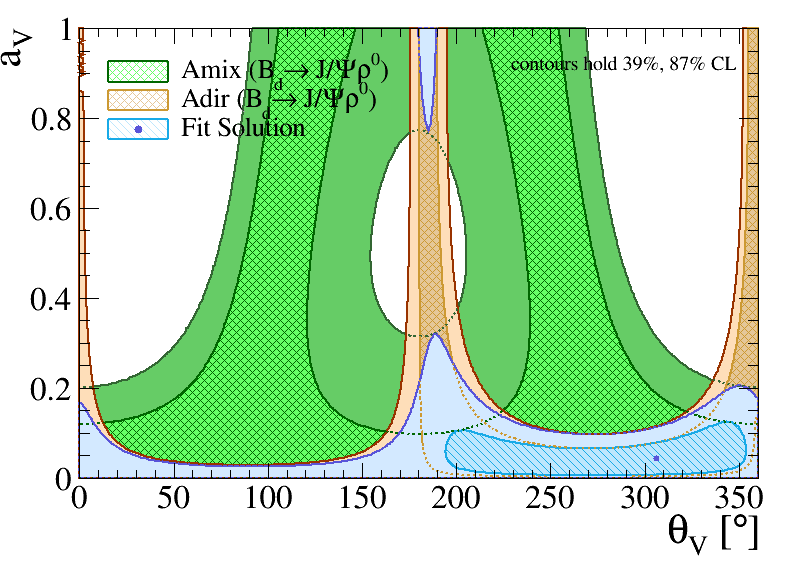}
    \includegraphics[width=0.49\textwidth]{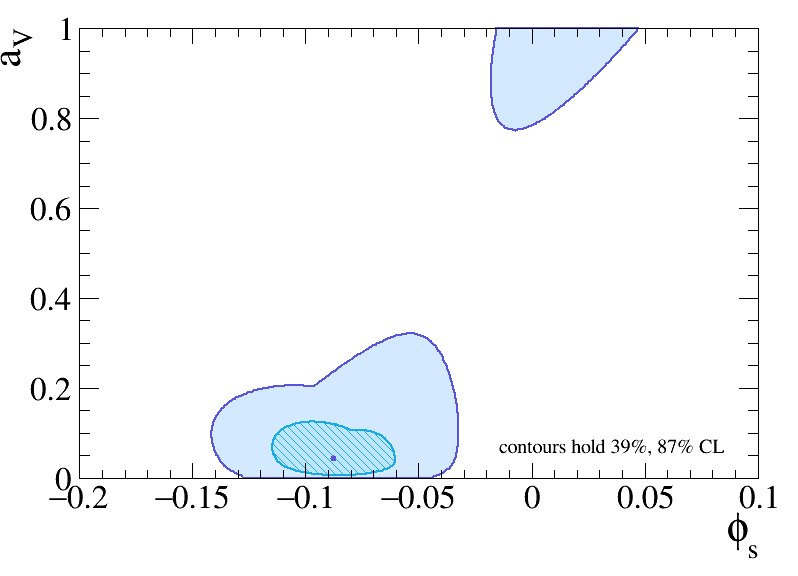}
    \caption{Two-dimensional confidence regions of the fit for the penguin parameters, $\phi_d$ and $\phi_s$ from the CP asymmetries in the \BJpsiX decays.
    Note that the contours for $\mathcal{A}_{\text{CP}}^{\text{dir}}$ and $\mathcal{A}_{\text{CP}}^{\text{mix}}$ are added for illustration only.
    They include the best fit solutions for $\phi_d$, $\phi_s$ and $\gamma$ as Gaussian constraints.
    }
    \label{fig:MasterFit}
\end{figure}

\begin{figure}
    \centering
    \includegraphics[width=0.49\textwidth]{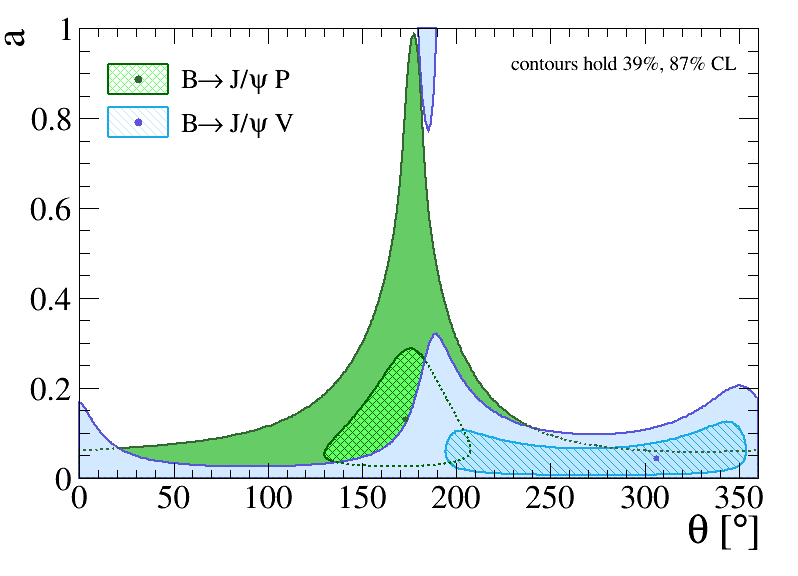}
    \caption{Comparison of the two-dimensional confidence regions of the fit solutions for the vector--pseudo-scalar and vector--vector final states.}
    \label{fig:MasterFit_JpsiP_JpsiV}
\end{figure}

To properly take this interplay between the \BJpsiX decay channels into account, we propose a combined fit to the CP asymmetries in the \BdJpsiPi \eqref{eq:Bd2JpsiPi}, \BdJpsiK \eqref{eq:Bd2JpsiK}, \BdJpsiRho \eqref{eq:Bd2JpsiRho}, \BsJpsiPhi \eqref{eq:BsJpsiPhi} and \BsJpsiKS \eqref{eq:BsJpsiK} modes, complemented with the average \eqref{eq:gamma} for $\gamma$ as an external constraint.
Utilising the $SU(3)$ symmetry relation \eqref{eq:SU3_pen}, we assume that the penguin parameters describing the \BdJpsiK, \BsJpsiKS and \BdJpsiPi channels are equal to one another, and similarly for the \BsJpsiPhi and \BdJpsiRho decays.
As justified in Section \ref{sec:Data_phi_d}, we will neglect possible $SU(3)$-breaking effects, given the lack of sensitivity in the current data, and we also neglect contributions from exchange and penguin-annihilation topologies.
For the vector--pseudo-scalar states, we obtain
\begin{equation}\label{eq:Master_VP}
    a = 0.13_{-0.10}^{+0.16}\:, \qquad
    \theta = \left(173_{-43}^{+34}\right)^{\circ}\:, \qquad
    \phi_d = \left(44.4_{-1.5}^{+1.6}\right)^{\circ}\:,
\end{equation}
and the solution for $a$ and $\theta$ corresponds to the phase shift
\begin{equation}
    \Delta\phi_d = \left(-0.73_{-0.91}^{+0.60}\right)^{\circ}\:.
\end{equation}
For the vector--vector final states we get 
\begin{equation}\label{eq:Master_VV}
    a_V = 0.043_{-0.037}^{+0.082}\:, \qquad
    \theta_V = \left(306_{-112}^{+\phantom{0}48}\right)^{\circ}\:, \qquad
    \phi_s = -0.088_{-0.027}^{+0.028} = \left(-5.0_{-1.5}^{+ 1.6}\right)^{\circ}\:,
\end{equation}
and the solution for $a_V$ and $\theta_V$ yields
\begin{equation}
    \Delta\phi_s = 0.003_{-0.012}^{+0.010} = \left(0.14_{-0.70}^{+0.54}\right)^{\circ}\:.
\end{equation}
The two-dimensional confidence regions of the simultaneous fit are shown in Fig.~\ref{fig:MasterFit}.
In comparison with Fig.~\ref{fig:Bd2JpsiX}, the second solution for $a$ and $\theta$ has disappeared due to the added constraints from the  CP asymmetries of the \BsJpsiKS decay.
Nonetheless, the strong correlation between $a$ and $\phi_d$ remains.
For the vector--vector final states, the correlation between $a_V$ and $\phi_s$ is a lot smaller.
Fig.~\ref{fig:MasterFit_JpsiP_JpsiV} shows a direct comparison between the fit solutions $(a,\theta)$, for the vector--pseudo-scalar, and $(a_V, \theta_V)$, for the vector--vector final states.
Although the results are still compatible with each other given the large uncertainties, the completely different shapes of the confidence regions illustrate the different decay dynamics of the vector--pseudo-scalar and vector--vector modes, which is expected on theoretical grounds.
It is therefore necessary to analyse different classes of final states independently, and we may in particular not assume $ae^{i\theta} = a_Ve^{i\theta_V}$.


The results for the mixing phases $\phi_d$ and $\phi_s$ in Eqs.\ \eqref{eq:Master_VP} and \eqref{eq:Master_VV} are corrected for possible contributions from penguin topologies, and represent the key findings of our analysis.
Comparing them to the SM predictions in Eqs.~\eqref{eq:phi_SM_E3}--\eqref{eq:phi_SM_I2} allows us to explore the space available for NP contributions:
\begin{align}
    \text{Excl,} K\ell3 & &
    \phi_d^{\text{NP}} & = (-1.3 \pm 2.6)^{\circ}\:, & 
    \phi_s^{\text{NP}} & = -0.050 \pm 0.028 = (-2.9 \pm 1.6)^{\circ}\:, \\
    \text{Excl,} K\ell2 & &
    \phi_d^{\text{NP}} & = (-0.8 \pm 2.4)^{\circ}\:, & 
    \phi_s^{\text{NP}} & = -0.050 \pm 0.027 = (-2.9 \pm 1.6)^{\circ}\:,\\
    \text{Incl,} K\ell3 & &
    \phi_d^{\text{NP}} & = (-8.3 \pm 2.8)^{\circ}\:, & 
    \phi_s^{\text{NP}} & = -0.045 \pm 0.028 = (-2.5 \pm 1.6)^{\circ}\:,\\
    \text{Incl,} K\ell2 & &
    \phi_d^{\text{NP}} & = (-7.7 \pm 2.8)^{\circ}\:, & 
    \phi_s^{\text{NP}} & = -0.044 \pm 0.028 = (-2.5 \pm 1.6)^{\circ}\:.
\end{align}
The picture emerging for $\phi_s^{\text{NP}}$ is consistent among the four SM scenarios, with a significance between 1.5 and 1.8 standard deviations.
In addition, the precision on this result is limited by the experimental fit \eqref{eq:Master_VV}, and will remain so for the foreseeable future.
Therefore, $\phi_s$ remains a powerful probe to search for NP effects and it will be interesting to see how this picture evolves over the coming years.
For $\phi_d$, the situation is very different.
The precision of $\phi_d^{\text{NP}}$ is already limited by the uncertainty of the SM prediction, and the significance strongly depends on the chosen SM scenario, varying from 0.3 to 3 standard deviations.
A resolution of the discrepancy between the inclusive and exclusive determinations of $|V_{ub}|$ and $|V_{cb}|$ is thus essential for NP searches using the $B_d^0$--$\bar B_d^0$ mixing phase.

%
%
%
\section{Hadronic Decay Benchmark Parameters}\label{sec:BR}
Let us now have a closer look at the information encoded in the branching fractions of the \BdJpsiK, \BdJpsiPi and \BsJpsiKS decays.
These quantities not only depend on the penguin parameters $a$ and $\theta$, but also on an overall normalisation factor.
Since we know the penguin parameters from the fit \eqref{eq:Master_VP} to the CP asymmetries, combining them with the experimental measurements of the branching fractions can give us valuable insights into very difficult to calculate hadronic parameters associated with the normalisation factor \eqref{eq:HadAmp_ccs}.
In particular, it allows us to determine a decay-specific effective colour-suppression factor.
The ratios of these factors between different decay channels provide insight into non-factorisable $SU(3)$ breaking effects.

Compared to the CP asymmetries, the normalisation factor is more sensitive to the chosen values for $\lambda$, $|V_{ub}|$ and $|V_{cb}|$.
For the discussion here, we will only illustrate the situation for one of the four scenarios introduced in Section \ref{sec:SM}, choosing the $K\ell3$ value \eqref{eq:Vus} for $|V_{us}|$, and the exclusive measurements \eqref{eq:Vb_excl} for $|V_{ub}|$ and $|V_{cb}|$.
For the other three scenarios, we can expect similar numerical variations as seen for $\phi_q^{\text{SM}}$ and $\phi_q^{\text{NP}}$.

\subsection{Decay Amplitudes and Branching Fractions}

The theoretical calculation of the decay amplitudes of the non-leptonic \BJpsiP decays is done using an effective field theory where all heavy degrees of freedom, i.e.\ the $W$ boson and top quark in the SM, are integrated out from appearing explicitly.
The transitions are described by a low-energy effective Hamiltonian
\begin{equation}
    A(\BJpsiP) = \langle J/\psi P| \mathcal{H}_{\text{eff}} | B_q^0\rangle\:,
\end{equation}
consisting of four-quark operators and their associated short-distance coefficients (for more background information, see,  for instance, Refs.~\cite{Buras:1985xv,Buchalla:1995vs,Buras:1998ra}).
The various local operators of the Hamiltonian represent the different decay topologies, such as tree and penguin contributions.

While the short-distance coefficients can be calculated within perturbation theory, the hadronic matrix elements of the four-quark operators require different tools and approximations.
A widely used approach in the literature ``factorises'' these matrix elements into the product of the hadronic matrix elements of the corresponding quark currents:
\begin{equation}\label{eq:ME_fact}
    \langle J/\psi P| (\bar c \gamma^{\mu} c) (\bar b \gamma_{\mu}q') | B_q^0\rangle|_{\text{fact}} = 
    \langle J/\psi| (\bar c \gamma^{\mu} c) | 0\rangle \langle P|(\bar b \gamma_{\mu}q') | B_q^0\rangle\:,
\end{equation}
where $\gamma_\mu$ are Dirac matrices, and $q'$ denotes a strange- or down-quark field.
The first term can be parametrised as
\begin{equation}\label{eq:ME_Jpsi}
  \langle 0|\bar c \gamma^{\mu} c| J/\psi\rangle = m_{J/\psi} f_{J/\psi} \varepsilon_{J/\psi}^{\mu}\:,
\end{equation}
where $m_{J/\psi}$ and $f_{J/\psi}$ are the mass and decay constant of the $J/\psi$ meson, respectively, and $\varepsilon_{J/\psi}$ is its polarisation vector.
The most recent lattice QCD calculation \cite{Hatton:2020qhk} gives 
\begin{equation}
    f_{J/\psi} = (410.4 \pm 1.7)\:\text{MeV}\:.
\end{equation}
The second matrix element 
\begin{align}\label{eq:FormFactor}
 \langle P|\bar b \gamma_{\mu}q'B\rangle = & \left[p_{B\mu} + p_{P\mu}-\left(\frac{m_B^2-m_P^2}{q^2}\right)q_{\mu}\right]f_{B\to P}^+(q^2) \nonumber \\
    & + \left(\frac{m_B^2-m_P^2}{q^2}\right)q_{\mu} \, f_{B\to P}^0(q^2)\:
\end{align}
can be parametrised in terms of hadronic $B\to P$ form factors $f_{B\to P}^0$ and $f_{B\to P}^+$, where $p_B$ and $p_P$ are the four momentum vectors of the corresponding mesons, and $q_{\mu} = p_{B\mu} - p_{P\mu}$ their momentum transfer.
Since the form factor $f_{B\to P}^0$ does not contribute to the product in \eqref{eq:ME_fact}, it does not affect the decays considered in this paper.

Let us for a moment consider only the two current--current operators
\begin{align}
    \mathcal{O}_1 & = \left(\bar c_\alpha \gamma_\mu(1-\gamma_5)q'_\beta\right)
    \left(\bar b_\beta \gamma^\mu(1-\gamma_5)c_\alpha\right)
    \equiv (\bar c_\alpha q'_\beta)_{\text{V}-\text{A}}(\bar b_\beta c_\alpha)_{\text{V}-\text{A}}\:,\\
    \mathcal{O}_2 & = \left(\bar c_\beta \gamma_\mu(1-\gamma_5)q'_\beta\right)
    \left(\bar b_\alpha \gamma^\mu(1-\gamma_5)c_\alpha\right)
    \equiv (\bar c_\beta q'_\beta)_{\text{V}-\text{A}}(\bar b_\alpha c_\alpha)_{\text{V}-\text{A}}\:,
\end{align}
where $\gamma_5$ is another Dirac matrix and $\alpha$, $\beta$ denote $SU(3)_{\text{C}}$ colour indices.
Making Fierz transformations and applying colour algebra relations, we obtain the standard expression for colour-suppressed (\emph{type-II}) $B$-meson decays in naive factorisation: 
\begin{equation}\label{eq:HadAmp_tree_fact} 
    A(\BJpsiP)|^{\text{tree}}_{\text{fact}} = \frac{G_{\text{F}}}{\sqrt{2}}V_{cq'}^{\phantom{*}}V_{cb}^*\, a_2\, 
     \langle J/\psi P| (\bar c \gamma^{\mu} c) (\bar b \gamma_{\mu}q) | B_q^0\rangle|_{\text{fact}} .
\end{equation}
Here $G_{\text{F}}$ is the Fermi constant, $V_{cq'}$ and $V_{cb}$ denote CKM matrix elements and the factorised matrix element is given in Eq.~\eqref{eq:ME_fact}.
It should be noted that the axial-vector components of the V$-$A operators do not contribute to the corresponding matrix elements of the quark currents as the $J/\psi$ is a vector meson and $P$ and $B_q^0$ are pseudo-scalar mesons (see \eqref{eq:ME_Jpsi} and \eqref{eq:FormFactor}).
The quantity 
\begin{equation}\label{eq:a2_wilson}
    a_2 = C_1 + \frac{C_2}{3}
\end{equation}
is a phenomenological ``colour suppression" factor, where $C_1$ and $C_2$ are the short-distance Wilson coefficients of the current--current operators $\mathcal{O}_1$ and  $\mathcal{O}_2$, respectively.
The naive parameter $a_2$ is typically found in the $0.1$--$0.3$ range \cite{Buras:1998us}.
Since these short-distance functions actually depend on the renormalisation scale $\mu$, while the $J/\psi$ decay constant and form factors entering \eqref{eq:ME_fact} do not depend on $\mu$, the factorised amplitude in \eqref{eq:HadAmp_tree_fact} depends on the renormalisation scale, which is unphysical.
This scale dependence is cancelled through non-factorisable contributions to the hadronic matrix elements of the four-quark operators $\mathcal{O}_1$ and $\mathcal{O}_2$, which cannot be calculated in a reliable way.
In order to circumvent this problem, a ``factorisation scale" $\mu_{\text{F}}$ is considered.
However, as $a_2$ depends strongly on $\mu$, we conclude that factorisation is not expected to work well for such colour-suppressed decays, as is well known in the literature.
Let us briefly note that in the case of colour-allowed (\emph{type-I}) decays, where the coefficient
\begin{equation}
    a_1 = \frac{C_1}{3} + C_2
\end{equation}
enters, the situation is more favourable.

The structure we obtained for the tree amplitude within the factorisation framework in Eq.~\eqref{eq:HadAmp_tree_fact} can be generalised to the full transition amplitude of the \BJpsiP decay, allowing also for penguin and non-factorisable effects:
\begin{align}\label{eq:DecAmp_fact}
    \sqrt{2}\: A(\BdJpsiPi) =
    & \frac{G_{\text{F}}}{\sqrt{2}}V_{cd}^{\phantom{*}}V_{cb}^*\: m_{J/\psi}\:
    f_{J/\psi} f_{B_d\to \pi}^+(m_{J/\psi}^2)\:
    (p_{B\mu}+p_{K\mu})\cdot\varepsilon_{J/\psi}^{\mu}\nonumber \\
    & \times \left(1- a e^{i\theta}e^{i\gamma}\right) \times a_2 (\BdJpsiPi)\:.
\end{align}
Here the $\sqrt{2}$ originates from the wave function of the neutral pion, $\pi^0 = (u\bar u + d\bar d)/\sqrt{2}$, and $a_2(\BdJpsiPi)$ is a generalisation of the naive colour-suppression factor in Eq.~\eqref{eq:a2_wilson}.
It is process-dependent and a renomalisation scale (and scheme) independent physical quantity which can be extracted from experimental data.
As can be seen in Eq.\ \eqref{eq:HadAmp_ccs}, it does not only get a contribution from the colour-suppressed tree amplitude but also from penguin topologies.

It would be very interesting to extract $a_2(\BdJpsiPi)$ in the cleanest possible way from experimental data, thereby shedding light on the importance of colour suppression and non-factorisable effects in \BJpsiP decays.
This is possible with the help of the branching fraction information.
Using Eq.~\eqref{eq:DecAmp_fact}, the CP-averaged  branching fraction of the \BdJpsiPi decay is given as
\begin{align}\label{eq:BRdef}
   2 \: \mathcal{B}(\BdJpsiPi) =  & 
   \: \tau_{B_d} \: \frac{G_{\mathrm F}^2}{32 \pi} |V_{cd}V_{cb}|^2 \: m_{B_d}^3
   \left[ f_{J/\psi}  f_{B_d\to \pi}^+(m_{J/\psi}^2) \right]^2 
    \left[\Phi\left(\frac{m_{J/\psi}}{m_{B_d}},\frac{m_{\pi^0}}{m_{B_d}}\right)\right]^3\nonumber \\
    & \times (1 - 2 a\cos\theta\cos\gamma + a^2) \times \left[ a_2 (\BdJpsiPi) \right]^2,
\end{align}
where the factor 2 on the left-hand side originates again from the $\pi^0$ wave function, $\tau_{B_d}$ is the lifetime of the $B_d^0$ meson, and 
\begin{equation}
    \Phi(x,y) = \sqrt{\left[1-(x+y)^2\right]\left[1-(x-y)^2\right]}
\end{equation} 
is the standard two-body phase-space function.
Similar expressions for the \BdJpsiK and \BsJpsiKS decays can be obtained by making straightforward substitutions.
Accurate knowledge of the penguin parameters is thus also needed to determine the effective colour-suppression factor $a_2$.

Note that for the interpretation of branching fraction measurements, subtleties arise due to effects originating from $B_q^0$--$\bar B_q^0$ mixing.
The experimentally measured ``time-integrated'' branching fraction is related to the ``theoretical" branching fraction given in Eq.~\eqref{eq:BRdef} by a correction factor \cite{DeBruyn:2012wj,DeBruyn:2012wk}, which depends on the decay width difference $\Delta\Gamma_q$ and the mass eigenstate rate asymmetry $\mathcal{A}_{\Delta\Gamma}(B_q\to f)$.
Since $\Delta\Gamma_d\approx 0$, this correction factor is 1 in the $B_d$-meson system with excellent precision.
On the other hand, for the $B_s$-meson system, the decay width difference is sizeable, thereby leading to corrections that can be as large as 10\%, depending on the final state \cite{DeBruyn:2012wj}.

\subsection{Form Factor Information}

The hadronic form factors have been calculated with a variety of approaches, most notably using lattice QCD \cite{Aoki:2019cca}.
The lattice results are obtained at high $q^2$ values and need to be extrapolated to the lower kinematic point $q^2 = m_{J/\psi}^2$ in order to use Eq.~\eqref{eq:BRdef} to determine the colour-suppression factor $a_2(\BdJpsiPi)$ from the data.
The extrapolation is typically done using the Bourrely--Caprini--Lellouch (BCL) parametrisation \cite{Bourrely:2008za}.
We obtain the following numerical values from the parameters provided by the FLAG \cite{Aoki:2019cca}:
\begin{align}
    f_{B_d\to\pi}^+(m_{J/\psi}^2) & = 0.371 \pm 0.069 \:,\label{eq:LFF_B2pi}\\
    f_{B_d\to K}^+(m_{J/\psi}^2) & = 0.645 \pm 0.022\:,\\
    f_{B_s\to K}^+(m_{J/\psi}^2) & = 0.470 \pm 0.024\:.\label{eq:LFF_Bs2K}
\end{align}
These results lead to the constraints
\begin{align}
    |a_2(\BdJpsiPi)|^2 \times (1-2 a \cos\theta\cos\gamma + a^2) & = 0.145 \pm 0.055\:, \\
    |a'_2(\BdJpsiK)|^2 \times (1+2\epsilon a\cos\theta\cos\gamma + \epsilon^2 a^2) & = 0.0714 \pm 0.0059\:, \\
    |a_2(\BsJpsiKS)|^2 \times (1-2 a \cos\theta\cos\gamma + a^2) & = 0.097 \pm 0.013\:,
\end{align}
which, in combination with the solution \eqref{eq:Master_VP} for the penguin parameters $a$ and $\theta$, yield
\begin{align}
    a_2(\BdJpsiPi) & = 0.363_{-0.079}^{+0.066}\:, \label{eq:a2_Bdpi_FF}\\
    a'_2(\BdJpsiK) & = 0.268_{-0.012}^{+0.011}\:,\label{eq:a2_B2K}\\
    a_2(\BsJpsiKS) & = 0.296_{-0.027}^{+0.024}\:.\label{eq:a2_Bs2K}
\end{align}
The results for the \BdJpsiK and \BsJpsiKS decays agree well with the theoretical estimates from naive factorisation.
The result for the \BdJpsiPi decay has a larger uncertainty in comparison with $a'_2(\BdJpsiK)$ and $a_2(\BsJpsiKS)$.
This can be traced back to the large uncertainty of the $B_d\to\pi$ form factor in Eq.~\eqref{eq:LFF_B2pi}.
It illustrates the current limitations of the lattice calculations, which are easier to compute and therefore more accurate when involving heavier particles.

\begin{figure}
    \centering
    \includegraphics[width=0.8\textwidth]{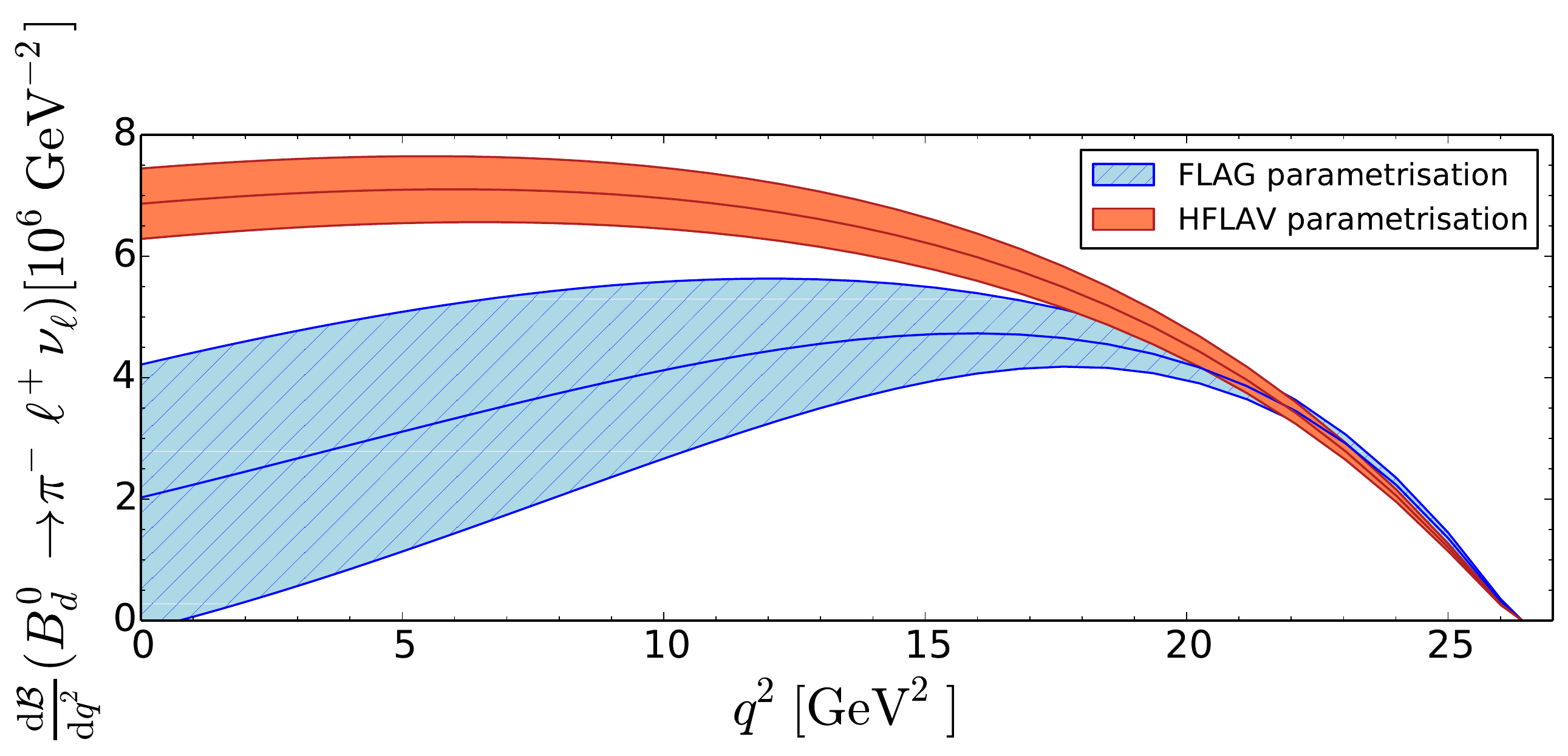}
    \caption{Comparison between the \Bdpilnu differential branching fraction distributions based on form factor parametrisations from FLAG (blue/hatched) and HFLAV (red).
    The uncertainty bands are due to the form factor parametrisations only.}
    \label{fig:form_factor}
\end{figure}

The limited precision of the available lattice calculations for the $B_d\to\pi$ form factor gets amplified in the extrapolation to lower $q^2$ values.
This becomes most apparent when looking at the differential branching fraction for the semileptonic \Bdpilnu decay.
In the limit $m_{\ell}\to 0$, this differential rate takes the form
\begin{equation}\label{eq:dGdq}
    \frac{d\Gamma}{dq^2}(\Bdpilnu) = \frac{G_{\text{F}}^2}{24\pi^3}|V_{ub}|^2\eta_{\text{EW}}^2p_{\pi}^3\left[f_{B_d\to\pi}^+(q^2)\right]^2\:,
\end{equation}
where $\eta_{\text{EW}} = 1.0066 \pm 0.0050$ \cite{Sirlin:1981ie} is the one-loop electroweak correction factor, and
\begin{equation}
    p_{\pi} = \frac{m_{B_d}}{2}\Phi\left(\frac{m_{\pi}}{m_{B_d}},\frac{q}{m_{B_d}}\right)
\end{equation}
is the pion momentum in the rest frame of the decaying $B_d$ meson.
Plotting the predicted rate for the \Bdpilnu differential branching fraction with the help of the FLAG parametrisation \cite{Aoki:2019cca}, which was also used to obtain Eq.~\eqref{eq:LFF_B2pi}, results in the blue curve shown in Fig.\ \ref{fig:form_factor}.

On the other hand, this rate has been measured experimentally by the BaBar and Belle collaborations.
HFLAV \cite{Amhis:2019ckw} has combined this information with lattice QCD and light cone sum rule (LCSR) calculations to provide an alternative set of BCL parameters for the $B_d\to\pi$ form factor.
This leads to the numerical result
\begin{equation}\label{eq:EFF_B2pi}
    f_{B_d\to\pi}^+(m_{J/\psi}^2) = 0.487 \pm 0.018\:,
\end{equation}
which is in much better agreement with the $B_s\to K$ form factor \eqref{eq:LFF_Bs2K}, as would be expected on the basis of $SU(3)$ flavour symmetry.
The predicted differential rate for the \Bdpilnu decay, which by construction matches the experimental data, is given by the red curve in Fig.\ \ref{fig:form_factor}.
We observe a large discrepancy between both curves, which demonstrates the challenges with extrapolating the lattice results and illustrates the theoretical uncertainty associated with the form factors.
It would therefore be advantageous if the form-factor information could be avoided as much as possible.

\subsection{Semileptonic Decay Information}

Interestingly, the \BdJpsiPi branching fraction in Eq.~\eqref{eq:BRdef} and the \Bdpilnu semileptonic differential decay rate \eqref{eq:dGdq} have the same form-factor dependence.
Consequently, the hadronic form factors cancel in the ratio
\begin{equation}\label{eq:SL_ratio}
    R_d^\pi \equiv \frac{\Gamma(\BdJpsiPi)}{d\Gamma/dq^2|_{q^2=m_{J/\psi}^2}(\Bdpilnu)}
    = \frac{\mathcal{B}(\BdJpsiPi)}{d\mathcal{B}/dq^2|_{q^2=m_{J/\psi}^2}(\Bdpilnu)}\:.
\end{equation}
Using the relation
\begin{equation}
    \left|\frac{V_{cd}V_{cb}}{V_{ub}}\right|^2 = \frac{1-\lambda^2+{\cal O}(\lambda^4)}{R_b^2},
\end{equation}
where $\lambda$ and $R_b$ are given in Eqs.\ \eqref{eq:Vus} and \eqref{eq:Rb_excl}, respectively, and neglecting the $\mathcal{O}(\lambda^4)$ corrections, we obtain 
\begin{equation}\label{eq:RdPi}
    R_d^\pi = 3\pi^2\left(\frac{1-\lambda^2}{R_b^2}\right)
   \left(\frac{f_{J/\psi}}{\eta_{\text{EW}}}\right)^2\times(1-2 a \cos\theta\cos\gamma + a^2) \times \left[ a_2 (\BdJpsiPi)\right]^2 \:.
\end{equation}
This expression allows us to determine the colour-suppression factor $a_2 (\BdJpsiPi)$ in a theoretically clean way that is not affected by form-factor uncertainties.
Knowledge of the penguin parameters is still required, though.

A similar ratio can be constructed for the \BsJpsiKS channel, using the semileptonic \BsKlnu decay.
It takes the form
\begin{equation}
    R_s^K \equiv \frac{\Gamma(\BsJpsiKS)}{d\Gamma/dq^2|_{q^2=m_{J/\psi}^2}(\BsKlnu)}
    = \frac{\mathcal{B}(\BsJpsiKS)}{d\mathcal{B}/dq^2|_{q^2=m_{J/\psi}^2}(\BsKlnu)}\:.
\end{equation}
The expression in terms of the \BsJpsiKS penguin parameters and colour-suppression factor is analogous to Eq.~\eqref{eq:RdPi}.
A first measurement of the \BsKlnu branching fraction was recently published by LHCb \cite{Aaij:2020nvo}.
However, we will have to wait for a future update that includes a measurement of the differential branching fraction to calculate $R_s^K$.
Until then, the hadronic form-factor information remains needed.
Finally, no semileptonic partner exists for the \BdJpsiK channel, and hadronic form-factor information will thus always be required to analyse this decay.

The $B\to\pi\ell\nu$ differential branching fraction has been measured by the BaBar and Belle experiments, which assume isospin symmetry to combine the experimental data from both \Bdpilnu and $B^+\to\pi^0\ell^+\nu$ channels.
We use the experimental average \cite{Amhis:2019ckw} for the $q^2$ bin
\begin{equation}
    \frac{d\mathcal{B}}{dq^2|_{q^2=[8,10]\:\text{GeV}^2}}(B\to\pi\ell\nu) = (6.44 \pm 0.43) \times 10^{-6}\:\text{GeV}^{-2}
\end{equation}
to represent the value at $q^2 = m_{J/\psi}^2$.
Combining this result with the \BdJpsiPi branching fraction \cite{Zyla:2020zbs}, we obtain
\begin{equation}
    R_d^{\pi} = (2.58 \pm 0.23)\times 10^{-6}\:\text{MeV}^2\:,
\end{equation}
leading to the constraint
\begin{equation}\label{eq:a2_semi_constraint}
    \left[ a_2 (\BdJpsiPi)\right]^2 \times (1-2 a \cos\theta\cos\gamma + a^2) = (0.0832 \pm 0.0079)\:.
\end{equation}

\begin{figure}
    \centering
    \includegraphics[width=0.49\textwidth]{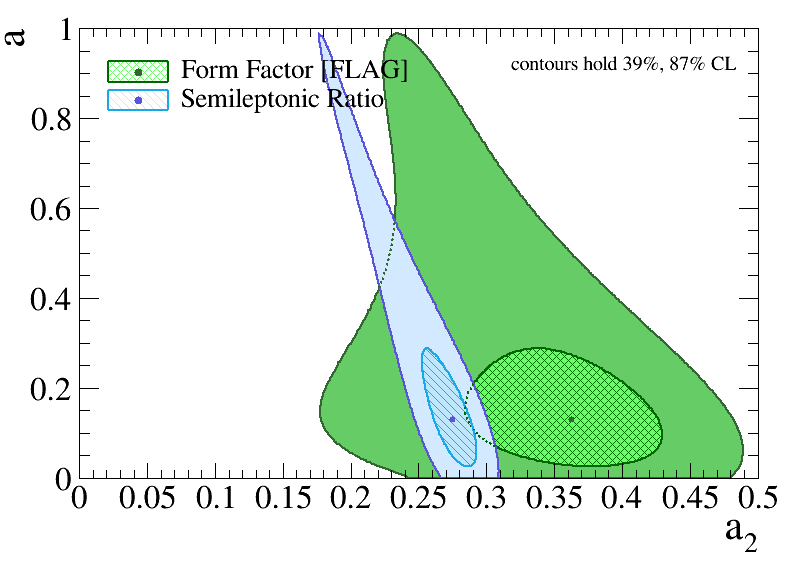}
    \caption{Comparison of the two-dimensional confidence regions of the fit for the colour-suppression factor $a_2(\BdJpsiPi)$ between the approach using form-factor input from lattice QCD calculations and the new method utilising the semileptonic ratio \eqref{eq:SL_ratio}.}
    \label{fig:SemiRat_a2}
\end{figure}

Adding the constraint \eqref{eq:a2_semi_constraint} to the GammaCombo fit \eqref{eq:Master_VP} gives
\begin{equation}\label{eq:a2_B2pi}
    a_2(\BdJpsiPi) = 0.275_{-0.023}^{+0.018}\:,
\end{equation}
which agrees much better with the theoretical estimates from naive factorisation than the form-factor based result in Eq.\ \eqref{eq:a2_Bdpi_FF}.
The correlation of the effective colour-suppression factor with the penguin parameter $a$ is given by the two-dimensional confidence regions shown in Fig.~\ref{fig:SemiRat_a2}.
In this figure, we also compare the result with the one obtained by using the lattice form factor parametrisation.
One immediately notices two things: the shift in the central value, and the much larger uncertainty.
The former is related to the discrepancy between the FLAG and HFLAV form factor parametrisations, as illustrated in Fig.~\ref{fig:form_factor}, while the latter is due to the large uncertainty of the form factor in Eq.~\eqref{eq:LFF_B2pi}.
This comparison demonstrates the advantages of our proposed strategy using the semileptonic ratio.

\begin{figure}
    \centering
    \includegraphics[width=0.49\textwidth]{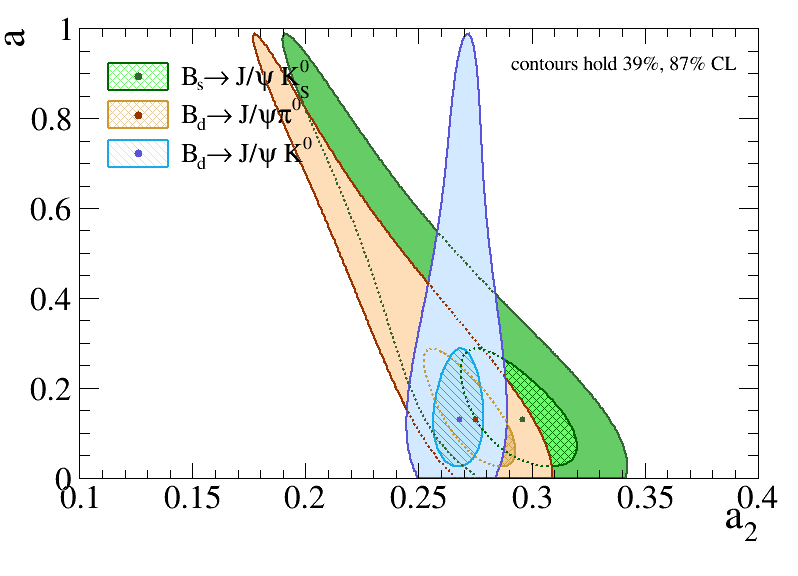}
    \includegraphics[width=0.49\textwidth]{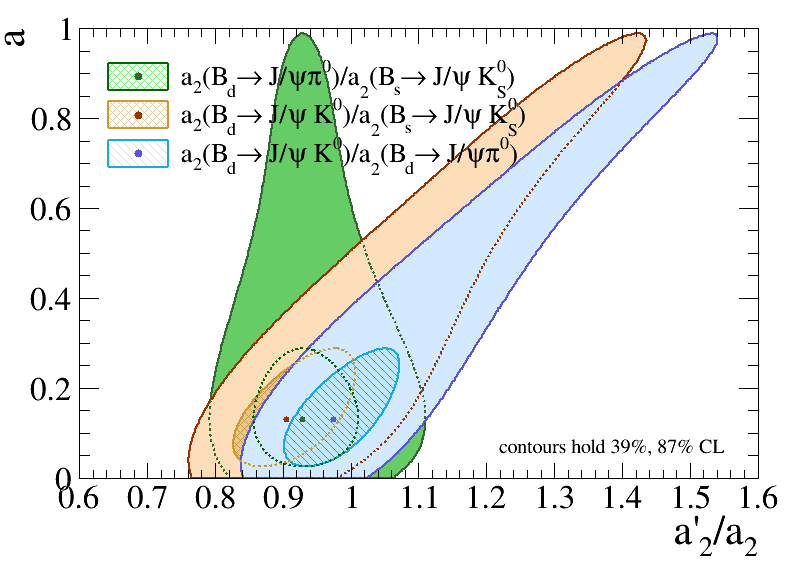}
    \caption{Two-dimensional confidence regions of the fit for the effective colour-suppression factors and their ratios.}
    \label{fig:a2_factor}
\end{figure}

Combining the results \eqref{eq:a2_B2pi}, \eqref{eq:a2_B2K} and \eqref{eq:a2_Bs2K}, we obtain the ratios
\begin{align}
    \frac{a'_2(\BdJpsiK)}{a_2(\BdJpsiPi)} & = 0.974_{-0.073}^{+0.098}\:, \label{eq:rat_a2_a}\\
    \frac{a'_2(\BdJpsiK)}{a_2(\BsJpsiKS)} & = 0.905_{-0.078}^{+0.101}\:, \label{eq:rat_a2_b}\\
    \frac{a_2(\BdJpsiPi)}{a_2(\BsJpsiKS)} & = 0.928_{-0.071}^{+0.083}\:. \label{eq:rat_a2_c}
\end{align}
The correlation of the effective colour-suppression factors and their ratios with the size $a$ of the penguin effects is given by the two-dimensional confidence regions shown in Fig.~\ref{fig:a2_factor}.
All three ratios are fully consistent with unity, as predicted in the strict limit of the $SU(3)$ flavour symmetry.
Consequently, they show that non-factorisable $SU(3)$-breaking effects are small, thereby supporting the assumptions we made in our analysis of the current data.
Fig.~\ref{fig:a2_factor} shows that the ratio $a'_2/a_2$ of colour-suppression factors is linearly correlated with the size $a$ of the penguin contributions.
Thus, also the size of non-factorisable $SU(3)$ breaking effects, i.e.\ the deviation of $a'_2/a_2$ from unity, is linearly correlated with $a$.
This means that if future experimental updates confirm the current picture for $a$, that the penguin effects are small, then also the non-factorisable $SU(3)$ breaking effects will correspondingly be small.

\subsection[About Non-Factorisable SU3 Breaking Corrections]{About Non-Factorisable $SU(3)$ Breaking Corrections}

It will be interesting to confront the results for the colour-suppression factor $a_2$ in Eqs.\ \eqref{eq:a2_B2K}, \eqref{eq:a2_Bs2K}, and \eqref{eq:a2_B2pi} with more sophisticated calculations within QCD factorisation or soft collinear effective theories.
It should be noted that they fall remarkably well into the $0.1$--$0.3$ range \cite{Buras:1998us} arising in naive factorisation, thereby not showing any anomalously large non-factorisable effects in the colour-suppressed tree topologies governing the \BJpsiP decays.
If we take the result $a_2 = 0.21 \pm 0.05$ quoted in Ref.\ \cite{Buras:1998us} as a reference, the values in Eqs.\ \eqref{eq:a2_B2K}, \eqref{eq:a2_Bs2K}, and \eqref{eq:a2_B2pi} leave room for deviations from factorisation at a level of 28\% to 41\%.
This is a non-trivial finding, since factorisation is --- a priori --- not expected to work well in these decays, and the non-factorisable effects could potentially have been much larger.

Lattice QCD \cite{Aoki:2019cca} and other non-perturbative methods \cite{Bracco:2011pg,El-Bennich:2016bno,Yao:2020vef} have illustrated that the flavour symmetries are broken.
The ratios between the kaon and pion decay constants, or between the $D_s$ and $D$ meson decay constants show that the $SU(3)$ flavour symmetry, which is relevant here, is generically broken at the 20\% level.
Consequently, putting both effects together, we expect non-factorisable $SU(3)$-breaking at the 5\%--8\% level.
This is actually confirmed by the experimental results in Eqs.\ \eqref{eq:rat_a2_a}, \eqref{eq:rat_a2_b}, and \eqref{eq:rat_a2_c}, showing non-factorisable $SU(3)$-breaking effects in the involved colour-suppressed tree topologies of at most $\mathcal{O}(10\%)$. 

These interesting results support the application of the $SU(3)$ flavour symmetry for the hadronic parameters $a$ and $\theta$, which are ratios of contributions of penguin topologies with respect to the colour-suppressed tree topologies.
In these ratios, the factorisable $SU(3)$-breaking effects, which are described by form factors and decay constants, cancel.
If we assume non-factorisable effects of up to 50\% due to the penguin topologies (i.e.\ much larger than for the colour-suppressed tree amplitudes) and again $SU(3)$-breaking effects at the 20\% level, this results in non-factorisable $SU(3)$-breaking effects in these quantities at the 10\% level, thereby illustrating the robustness of our strategy to such effects.
In the future, more precise data will not only result in a sharper picture for the penguin effects, but for the $SU(3)$-breaking effects as well.

%
%
%
\section{Future Perspectives}\label{sec:bench}
Although the current uncertainties of the penguin parameters and their impact on the mixing phases $\phi_q$ are still large, significant improvements can be expected over the next decades thanks to the large increase in luminosity expected from the HL-LHC \cite{Cerri:2018ypt} and Belle II \cite{Kou:2018nap} programmes.
In order to demonstrate the potential of the strategies proposed in this paper, we consider two benchmark scenarios for the combined fit of the five \BJpsiX decays presented in Subsection~\ref{sec:BJpsiX_fit}:
we will keep the central values of the current experimental results to allow for an easy comparison but divide all uncertainties by a factor 2 (first scenario), and by a factor 5 (second scenario).
These scenarios are chosen to demonstrate the potential of the strategy proposed in this paper, and do not take into account the different experimental challenges of the various decay channels, or the time scales on which these improvements can be achieved.
More detailed and sophisticated studies building upon the flagship analysis for the high-luminosity era of quark-flavour physics \cite{Bediaga:2018lhg,Kou:2018nap,Cerri:2018ypt} would be very desirable.

\begin{figure}
    \centering
    \includegraphics[width=0.49\textwidth]{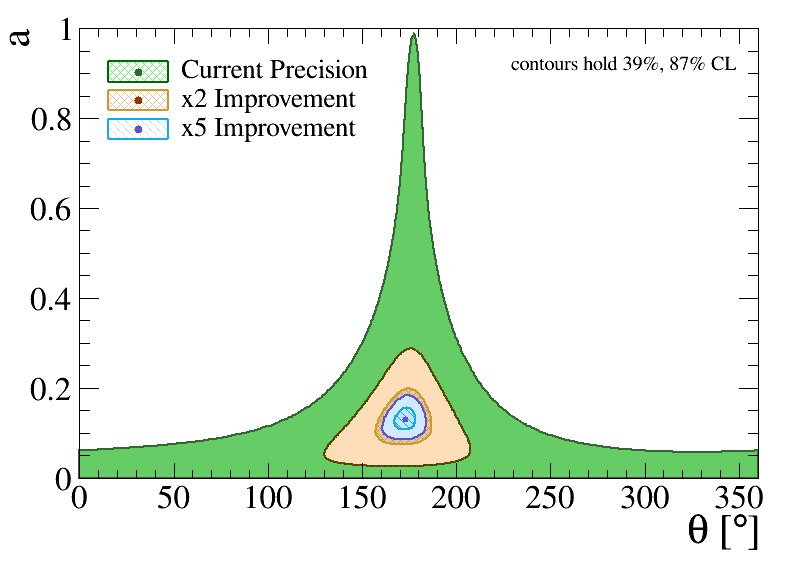}
    \includegraphics[width=0.49\textwidth]{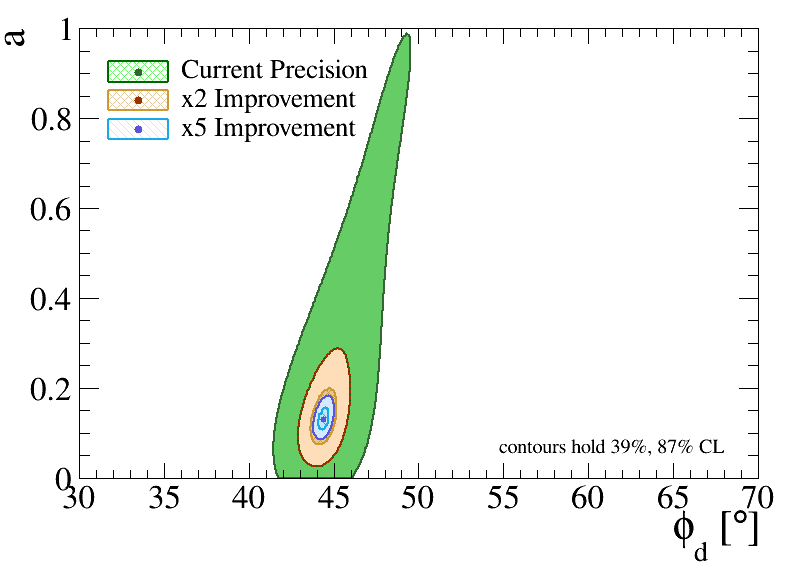}
    
    \includegraphics[width=0.49\textwidth]{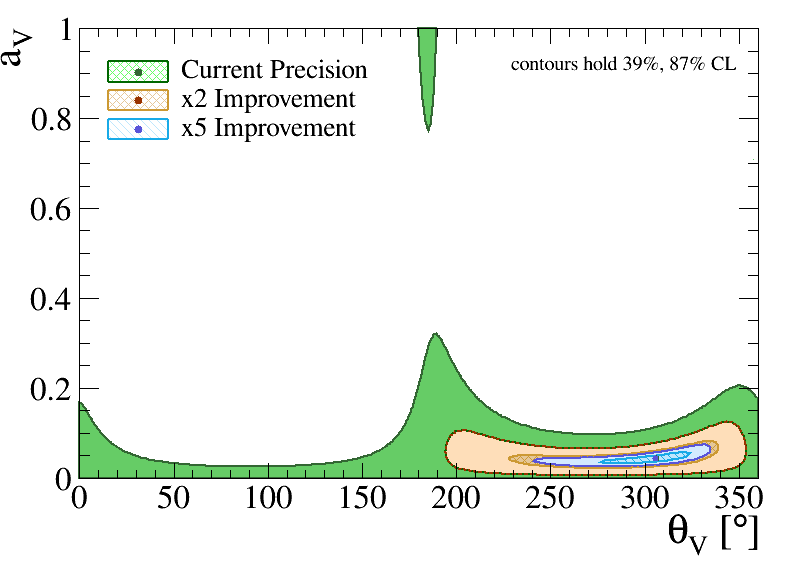}
    \includegraphics[width=0.49\textwidth]{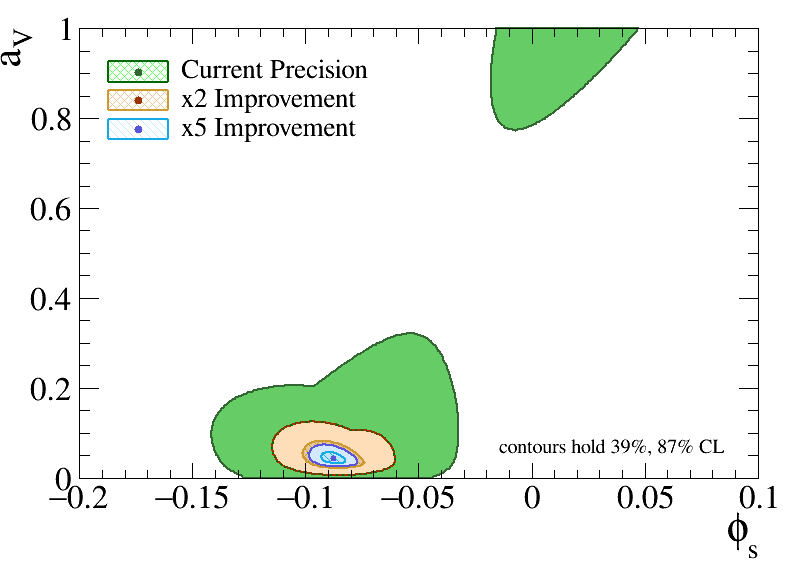}
    \caption{Comparison between the current precision of the penguin parameters and mixing phases with two future benchmark scenarios in which we assume an overall improvement of the experimental input measurements by a factor 2 and 5.
    Shown are the two-dimensional confidence regions of the fit for the penguin parameters, $\phi_d$ and $\phi_s$ from the CP asymmetries in the \BJpsiX decays.
    }
    \label{fig:future_benchmarks}
\end{figure}

\begin{figure}
    \centering
    \includegraphics[width=0.49\textwidth]{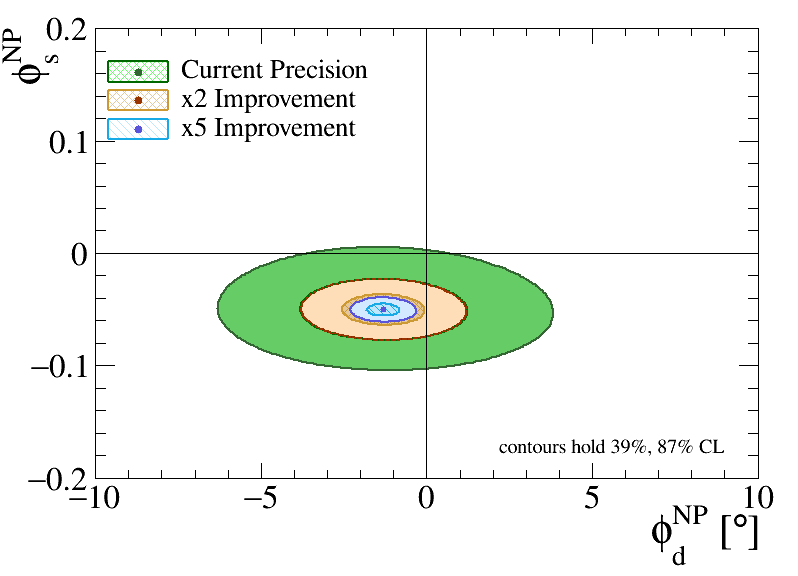}
    \caption{Comparison between the current precision of the phases $\phi_d^{\text{NP}}$ and $\phi_s^{\text{NP}}$ with two future benchmark scenarios in which we assume an overall improvement of the experimental input measurements by a factor 2 and 5.
    Shown are the two-dimensional confidence regions of the fit.}
    \label{fig:future_NP}
\end{figure}

The numerical results for the two benchmark scenarios are compared with the fit to the current data in Table \ref{tab:future_benchmarks}. 
The evolution of the two-dimensional confidence regions for the penguin parameters and the mixing phases are shown in Fig.~\ref{fig:future_benchmarks}, while the corresponding regions for the phases $\phi_d^{\text{NP}}$ and $\phi_s^{\text{NP}}$, based on the SM prediction \eqref{eq:phi_SM_E3}, is shown in Fig.\ \ref{fig:future_NP}.
We observe that already a factor 2 improvement in the experimental precision would have a large impact on the determination of the penguin parameters and would allow us to firmly establish non-zero penguin contributions.
We therefore advocate to give the CP asymmetry measurements of the penguin control modes the same importance as those of the flagship decays \BdJpsiK and \BsJpsiPhi.
This will prevent the penguin effects from becoming the dominant uncertainty in the determination of $\phi_d$ and $\phi_s$.

Imagine a situation where we reduce the experimental uncertainty of the CP asymmetry measurements of \BdJpsiK by a factor two, but no such improved measurements are yet available for the control modes \BdJpsiPi and \BsJpsiKS.
We will then know the effective mixing phase $\phi_d^{\text{eff}}$ with a precision of $0.7^{\circ}$, which would be matched by the precision of the penguin shift $\Delta\phi_d$ (see Table \ref{tab:future_benchmarks}).
This will then lead to a precision of $\phi_d$ of $1^{\circ}$.
When the experimental uncertainties on the CP asymmetry measurements of \BdJpsiPi and \BsJpsiKS also improve by a factor two, the precision of the penguin shift $\Delta\phi_d$ improves to $0.35^{\circ}$ and the precision of $\phi_d$ becomes $0.78^{\circ}$.
After updating the measurements of \BdJpsiK, \BdJpsiPi and \BsJpsiKS we thus know $\phi_d$ 22\% more precise than when only focusing on \BdJpsiK.
Adequately controlling the penguin contributions can thus have a large impact on our knowledge of $\phi_d$ and $\phi_s$.

The benchmark scenarios in Table \ref{tab:future_benchmarks} also illustrate that even though NP contributions to $\phi_s$ are small, it remains possible to control the penguin contributions with sufficient precision to establish non-zero values for $\phi_s^{\text{NP}}$ with a significance of more than five standard deviations.
For $\phi_d$ the situation is less clear.
Searches for NP contributions to $\phi_d$ are limited due to the unresolved discrepancies between the available SM predictions, as illustrated and discussed at the end of Section \ref{sec:CurrData}.
In Table \ref{tab:future_benchmarks} we have assumed the SM value \eqref{eq:phi_SM_E3} for all three scenarios.

\begin{table}
    \centering
    \begin{tabular}{|c|c|c|c|c|}
        \toprule
        Obs. & Best Fit & Current Precision & $\times 2$ Improvement & $\times 5$ Improvement \\
        \midrule
        $a$ & $0.131$ & $-0.10/+0.16$ & $-0.056/+0.068$ & $-0.024/+0.026$ \\
        $\theta$ & $172.8^{\circ}$ & $-43^{\circ}/+34^{\circ}$ & $-16^{\circ}/+14^{\circ}$ & $-5.8^{\circ}/+5.6^{\circ}$ \\
        $a_V$ & $0.043$ & $-0.037/+0.082$ & $-0.021/+0.039$ & $-0.011/+0.014$ \\
        $\theta_V$ & $306^{\circ}$ & $-112^{\circ}/+48^{\circ}$ & $-78^{\circ}/+33^{\circ}$ & $-31^{\circ}/+18^{\circ}$ \\
        \midrule
        $\phi_d$ & $44.38^{\circ}$ & $-1.5^{\circ}/+1.6^{\circ}$ & $-0.76^{\circ}/+0.79^{\circ}$ & $-0.31^{\circ}/+0.32^{\circ}$ \\
        $\Delta\phi_d$ & $-0.73^{\circ}$ & $-0.91^{\circ}/+0.60^{\circ}$ & $-0.40^{\circ}/+0.31^{\circ}$ & $-0.16^{\circ}/+0.13^{\circ}$ \\
        $\phi_d^{\text{SM}}$ & $45.67^{\circ}$ & $-2.0^{\circ}/+1.9^{\circ}$ & $-0.98^{\circ}/+0.98^{\circ}$ & $-0.39^{\circ}/+0.39^{\circ}$ \\
        $\phi_d^{\text{NP}}$ & $-1.29^{\circ}$ & $-2.4^{\circ}/+2.6^{\circ}$ & $-1.2^{\circ}/+1.3^{\circ}$ & $-0.50^{\circ}/+0.50^{\circ}$ \\
        \midrule
        $\phi_s$ & $-5.03^{\circ}$ & $-1.5^{\circ}/+1.6^{\circ}$ & $-0.74^{\circ}/+0.80^{\circ}$ & $-0.32^{\circ}/+0.32^{\circ}$ \\
        $\Delta\phi_s$ & $0.14^{\circ}$ & $-0.70^{\circ}/+0.54^{\circ}$ & $-0.32^{\circ}/+0.28^{\circ}$ & $-0.12^{\circ}/+0.11^{\circ}$ \\
        $\phi_s^{\text{SM}}$ & $-2.153^{\circ}$ & $-0.11^{\circ}/+0.11^{\circ}$ & $-0.057^{\circ}/+0.057^{\circ}$ & $-0.027^{\circ}/+0.027^{\circ}$ \\
        $\phi_s^{\text{NP}}$ & $-2.88^{\circ}$ & $-1.6^{\circ}/+1.6^{\circ}$ & $-0.75^{\circ}/+0.80^{\circ}$ & $-0.31^{\circ}/+0.32^{\circ}$ \\
        \midrule
        $\phi_s$ & $-0.0878$ & $-0.027/+0.028$ & $-0.013/+0.014$ & $-0.0055/+0.0055$ \\
        $\Delta\phi_s$ & $0.0025$ & $-0.012/+0.010$ & $-0.0056/+0.0049$ & $-0.0021/+0.0019$ \\
        $\phi_s^{\text{SM}}$ & $-0.03757$ & $-0.0019/+0.0020$ & $-0.0010/+0.0010$ & $-0.00047/+0.00046$ \\
        $\phi_s^{\text{NP}}$ & $-0.0502$ & $-0.027/+0.028$ & $-0.013/+0.014$ & $-0.0054/+0.0055$ \\
        \bottomrule
    \end{tabular}    
    \caption{Comparison of the current precision for the penguin parameters and mixing phases with two future scenarios that assume overall improvements of the experimental input measurements by factors 2 and 5, respectively.}
    \label{tab:future_benchmarks}
\end{table}

\begin{figure}
    \centering
    \includegraphics[width=0.49\textwidth]{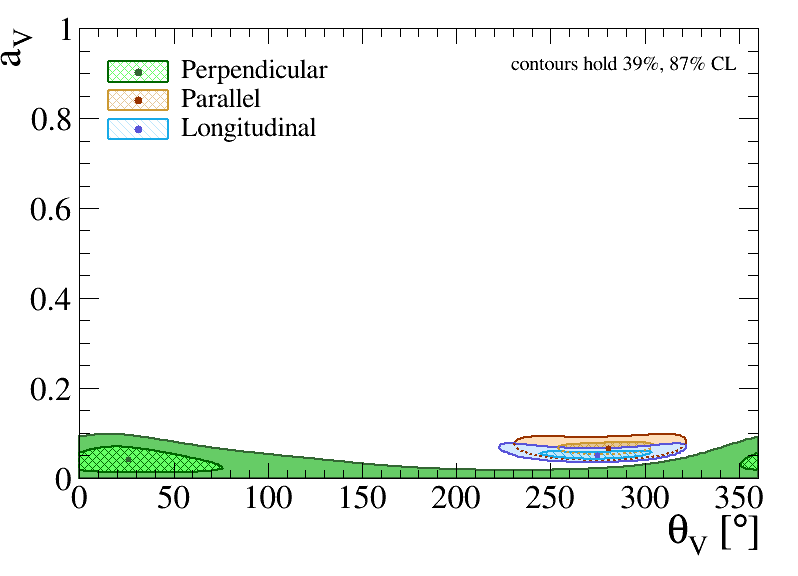}
    \caption{Comparison of the two-dimensional confidence regions of the fit for the penguin parameters from the CP asymmetries in \BdJpsiRho between the three polarisation states for the future benchmark scenario in which we assume an overall improvement of the experimental input measurements by a factor 5.
    (To be compared with Fig.~\ref{fig:Bd2JpsiRho}.)}
    \label{fig:future_benchmarks_poldep}
\end{figure}

These future benchmark scenarios also allow us to demonstrate the impact of polari\-sa\-tion-dependent measurements using the \BdJpsiRho decay as an example.
Reducing the uncertainties of the polarisation-dependent CP asymmetries used to obtain the confidence regions shown in Fig.~\ref{fig:Bd2JpsiRho} by a factor 5  would lead to the two-dimensional confidence regions shown in Fig.~\ref{fig:future_benchmarks_poldep}.
In this scenario, we would find a clear difference between the perpendicular polarisation on the one hand and the longitudinal and parallel polarisations on the other hand.
The polarisation-dependent penguin parameters would no longer be compatible with one another, leading to different penguin shifts $\Delta\phi_s^f$, given in Table \ref{tab:future_benchmarks_poldep}.

\begin{table}
    \centering
    \begin{tabular}{|c|r|c|c|c|}
        \toprule
        Obs. & Best Fit & Current Precision & $\times 2$ Improvement & $\times 5$ Improvement \\
        \midrule
        $\Delta\phi_s$ & $0.14^{\circ}$ & $-0.70^{\circ}/+0.54^{\circ}$ & $-0.32^{\circ}/+0.28^{\circ}$ & $-0.12^{\circ}/+0.11^{\circ}$ \\
        $\Delta\phi_s^0$ & $0.02^{\circ}$ & $-0.78^{\circ}/+0.58^{\circ}$ & $-0.36^{\circ}/+0.31^{\circ}$ & $-0.16^{\circ}/+0.16^{\circ}$ \\
        $\Delta\phi_s^{\parallel}$ & $0.07^{\circ}$ & $-0.91^{\circ}/+0.65^{\circ}$ & $-0.41^{\circ}/+0.35^{\circ}$ & $-0.18^{\circ}/+0.17^{\circ}$ \\
        $\Delta\phi_s^{\perp}$ & $0.21^{\circ}$ & $-0.91^{\circ}/+0.66^{\circ}$ & $-0.41^{\circ}/+0.36^{\circ}$ & $-0.18^{\circ}/+0.17^{\circ}$ \\
        \bottomrule
    \end{tabular}    
    \caption{Numerical comparison of the current precision for the polarisation-dependent penguin shifts with two future benchmark scenarios assuming overall improvements of the experimental input measurements by factors 2 and 5.}
    \label{tab:future_benchmarks_poldep}
\end{table}

\clearpage
%
%
%
\section{Conclusion}\label{sec:conclusion}
We have presented a state-of-the-art analysis of the penguin effects limiting the theoretical precision of the mixing phases $\phi_d$ and $\phi_s$ determined from \BdJpsiK and \BsJpsiPhi, respectively.
The corresponding control channels, utilising the $SU(3)$ flavour symmetry, are the \BdJpsiPi, \BsJpsiKS and \BdJpsiRho modes.
As the mixing-induced CP asymmetries of these five \BJpsiX channels also depend on $\phi_q$, we propose a simultaneous analysis of these decays, allowing us to extract the relevant hadronic parameters and the mixing phases
\begin{equation}
        \phi_d = \left(44.4_{-1.5}^{+1.6}\right)^{\circ}\:,\qquad
        \phi_s = -0.088_{-0.027}^{+0.028} = \left(-5.0_{-1.5}^{+ 1.6}\right)^{\circ}\:,
\end{equation}
directly taking the penguin effects into account.
These results can be averaged with the penguin-corrected measurements from other decay modes to further improve the experimental precision.
In the future, improvements will be possible once polarisation-dependent measurements of $\phi_s$ from \BsJpsiPhi become available.
Using future scenarios, we have shown that non-zero NP contributions to $\phi_s$ could be established with more than five standard deviations in the ultra high-precision era of flavour physics, which we can look forward to towards the end of the LHC and SuperKEKB upgrade programmes.
In this respect, the control of the penguin uncertainties is essential.
Consequently, we advocate to promote the corresponding control channels to benchmark decays for the exploration of CP violation.

Concerning the mixing phase $\phi_d$, we pointed out that the limiting factor for revealing NP effects is given by the SM prediction of $\beta$, which is governed by the UT side $R_b$.
For the current data, the SM uncertainty of about $2^{\circ}$ is already significantly larger than the experimental precision of $1.5^{\circ}$, taking the penguin effects into account.
In the future, it will hence be essential to improve the precision on $R_b$ and settle the tension between the inclusive and exclusive determinations of $|V_{ub}|$ and $|V_{cb}|$ from semileptonic $B$ decays.

Another important aspect of our analysis is the determination of the effective colour-suppression factors of the \BJpsiP decays, which serve as reference for future QCD calculations to compare against.
In the case of the \BdJpsiPi and \BsJpsiKS modes, we have proposed a new method using semileptonic \Bdpilnu and \BsKlnu decays, respectively, which does not require any information on hadronic form factors and is theoretically clean.
Unfortunately, data for the semileptonic $B_s^0$ decay are not yet available.
We have used lattice QCD results for form factors in this case and for the \BdJpsiK decay, where there is no semileptonic counterpart.
We obtain values of the effective colour-suppression factors in the ball park of theoretical expectations, which suffer from large uncertainties.
Furthermore, we have explored non-factorisable $SU(3)$-breaking effects in these quantities.
Interestingly, we could not reveal deviations from the $SU(3)$ limit within the current precision, which has already reached a level below 10\%.
This feature supports the use of the $SU(3)$ flavour symmetry to control the penguin parameters.
It will also be interesting to see further theoretical progress in the understanding of these hadronic parameters from first principles of QCD.

The strategies and decays discussed in this paper will play a key role for the long-term ultra high-precision flavour physics programme, offering exciting prospects to finally establish new sources of CP violation.

%
%
%
%
%
%
\section*{Acknowledgements}
We would like to thank Philine van Vliet for useful discussions.
This research has been supported by the Netherlands Organisation for Scientific Research (NWO).

\phantomsection 
\addcontentsline{toc}{section}{References}
\setboolean{inbibliography}{true}
\bibliographystyle{LHCb}
\bibliography{references}

\end{document}